\documentclass[10pt]{article}
\pdfoutput=1

\usepackage{ae}
\usepackage[T1]{fontenc}
\usepackage[ansinew]{inputenc}
\usepackage{mathrsfs}
\usepackage{amsmath}
\usepackage{amssymb}
\usepackage{dsfont}
\usepackage{graphicx}
\usepackage{upgreek}
\usepackage{color}
\usepackage{pst-all}
\usepackage{tikz}
\usepackage{import}
\usepackage{hyperref}
\usepackage{cite}

\usepackage{epsfig}
\usepackage{wasysym}

\usepackage{verbatim}
\usepackage{stmaryrd}

\usepackage{amsfonts}

\usepackage{tikz}
\usetikzlibrary{arrows,shapes}
\usetikzlibrary{patterns,fadings}

\usetikzlibrary{snakes,arrows,shapes,positioning,decorations}

\usepackage[a4paper, total={6.5in, 9.3in}]{geometry}
\usepackage{dsfont}
\usepackage{empheq}
\usepackage{slashed}
\usepackage{subcaption}
\usepackage{multirow}
\allowdisplaybreaks

\tikzset{
BPSbox/.style={
       fill={rgb,255: red,221; green,221; blue,221},
       draw={rgb,255: red,113; green,113; blue,113}, thick,
       text width=2.2cm,
       inner sep=2pt,
       text centered,
       }}
\tikzset{ BPSbox2/.style={
       fill={rgb,255: red,221; green,221; blue,221},
       draw={rgb,255: red,113; green,113; blue,113}, thick,
       text width=2.68cm,
       inner sep=2pt,
       text centered,
       }
}

\begin{document}

\thispagestyle{empty}

\renewcommand{\thefootnote}{\fnsymbol{footnote}}
\setcounter{footnote}{0}
\setcounter{figure}{0}
\begin{center}

{\Large
\textbf{\mathversion{bold}
On the abundance of supersymmetric strings in $AdS_3\times S^3\times S^3\times S^1$ describing BPS line operators
}}

\vspace{1.0cm}

\textrm{Diego H. Correa, Victor I. Giraldo-Rivera and Mart\'in Lagares  }
\\ \vspace{1.2cm}
\footnotesize{

\textit{ Instituto de F\'isica La Plata - CONICET \&\\
Departamento de F\'\i sica,  Universidad Nacional de La Plata\\ C.C. 67, 1900,  La Plata, Argentina} }

\vspace{3mm}

\par\vspace{1.5cm}

\textbf{Abstract}\vspace{2mm}
\end{center}

\noindent

We study supersymmetric open strings in type IIB $AdS_3 \times S^3 \times S^3 \times S^1$ with mixed R-R and  NS-NS fields. We focus on strings ending along a straight line at the boundary of $AdS_3$, which can be interpreted as line operators in a dual CFT$_2$. We study both classical configurations and quadratic  fluctuations around them. We find that strings sitting at a fixed point in $S^3 \times S^3 \times S^1$, i.e. satisfying Dirichlet boundary conditions, are 1/2 BPS. We also show that strings sitting at different points of certain submanifolds of $S^3 \times S^3 \times S^1$ can still share some fraction of the supersymmetry. This allows to define supersymmetric smeared configurations by the superposition of them, which range from 1/2 BPS to 1/8 BPS. In addition to the smeared configurations, there are as well
1/4 BPS and 1/8 BPS strings satisfying Neumann boundary conditions. 
All these supersymmetric strings are shown to be connected by a network of interpolating BPS boundary conditions.
Our study reveals the existence of a rich moduli of supersymmetric open string configurations, for which  the appearance of massless fermionic fields in the spectrum of quadratic fluctuations is crucial.

\vspace*{\fill}

\setcounter{page}{1}
\renewcommand{\thefootnote}{\arabic{footnote}}
\setcounter{footnote}{0}

\newpage
\tableofcontents

\section{Introduction}

Line operators, as for example Wilson loops, are perhaps the most basic non-local operators in any QFT, and they are central to the description of gauge field theories. Within the class of line operators, supersymmetric ones stand out, as their expectation values and correlation functions
can be described exactly in some cases \cite{Pestun:2007rz}.  Of special interest are the cases in which the lines also preserve the group of 1-dimensional conformal symmetries, as they would constitute superconformal defects. Remarkably, these two features have enabled a wide variety of non-trivial precision tests of the AdS/CFT correspondence, highlighting the importance of BPS line operators within this context.

The literature on supersymmetric line operators, especially for superconformal theories in $d=4$ and $d=3$ spacetime dimensions,  is  vast. As prototypical examples, we could mention Wilson loops in ${\cal N}=4$ super Yang-Mills \cite{Maldacena:1998im} and ${\cal N}=6$ super Chern-Simons-matter \cite{Drukker:2008zx,Chen:2008bp,Rey:2008bh,Drukker:2009hy}, whose dual descriptions are in terms of open strings in $AdS_5\times S^5$ and $AdS_4\times CP^3$ respectively (see \cite{Drukker:2019bev} for a review on Wilson loops in Chern-Simons-matter theories). There is however a striking difference between these cases that should be noted. While supersymmetric Wilson loops are pretty much fixed by the contour in ${\cal N}=4$ super Yang-Mills, one discovers a richer family in super Chern-Simons theories, as for a given contour it is possible to define families of supersymmetric Wilson loops containing arbitrary parameters \cite{Ouyang:2015iza,Ouyang:2015bmy}. The reasons why this is the case might not be so obvious in the field theory description. 
However, from the holographic dual point of view there is a hallmark  for this richness in the moduli of supersymmetric line operators: for the dual open strings on $AdS_4\times CP^3$ other boundary conditions than Dirichlet can also be consistent with supersymmetry  \cite{Correa:2019rdk}.

The fact that superconformal field theories in $d=3$ admits a richer moduli of superconformal lines seems to be a general property, irrespective of the aforementioned examples. In the article \cite{Agmon:2020pde}, a classification of superconformal line defects, within superconformal field theories in various spacetime dimensions ($3\leq d\leq 6$), has been given. This reveals a rich structure of marginal and relevant deformations for lines in $d=3$ superconformal theories, which are not admitted  for $d>3$. Recently, an extensive study of BPS Wilson loops in 3-dimensional supersymmetric gauge theories has been presented in \cite{Mauri:2017whf,Mauri:2018fsf,Drukker:2020opf,Drukker:2020dvr}.

It is natural then to ask whether this richness of supersymmetric line operators is available in other AdS/CFT setups. The purpose of this paper is to explore this question in the type IIB background $AdS_3\times S^3\times S^3\times S^1$. This is an interesting problem, as the holographic dual of the string theory on this background in the general case is more difficult to describe than in the cases in which the factor $S^3\times S^1$  is replaced either by $T^4$ or $K3$ \cite{Elitzur:1998mm,deBoer:1999gea,Gukov:2004ym,Tong:2014yna,Eberhardt:2017pty}, even more if one considers a non-vanishing RR 3-form flux. 
The study of superconformal line defects and the supersymmetries they preserve can set a basis for the description of the CFT$_2$ ambient theory allocating them. Superconformal symmetry constrains the correlation functions on the line, and physical quantities of the CFT$_2$ (such as the central charge or the Bremsstrahlung functions) could be inferred from them \cite{Gorini:2020new,Correa:2012at}.

More concretely, we study open strings ending along a straight line at the boundary of $AdS_3$ in quest of supersymmetric boundary conditions.
Our analysis begins with the simplest open string ending along a line while sitting at a point in the compact factors $S^3\times S^3\times S^1$, which turns out to be 1/2 BPS. Then, we move to consider open string configurations delocalized within some regions of $S^3\times S^3\times S^1$, but still preserving some fraction of the supersymmetry. We do this by smearing open strings,  satisfying Dirichlet boundary conditions for different values of the angular positions in $S^3\times S^3\times S^1$. These results are presented in section \ref{BPS classical strings}. 

More general delocalized string configurations, including the ones mentioned in the preceding paragraph, can be accounted in terms of the boundary conditions that are imposed on the quadratic fluctuations around the simplest 1/2 BPS open string. Fluctuations for angular positions in $S^3\times S^3\times S^1$ are described in terms of massless scalar fields in $AdS_2$, whose asymptotic behaviours are of the form
\begin{equation}
\phi^m(\tau,\sigma) = (\alpha^m(\tau)+\dots)+ \sigma(\beta^m(\tau)+\dots)\,,
\end{equation}
where $\sigma\to 0$ is the boundary of $AdS_2$. Since the vanishing mass is inside the Breitenlohner-Freedman window \cite{Breitenlohner:1982jf,Breitenlohner:1982bm}, it is possible to set either Dirichlet $\alpha^m = 0$ or 
Neumann $\beta^m = 0$ boundary conditions. In conjunction with massless spinor fields, boundary conditions for these scalars more general than Dirichlet can be supersymmetric \cite{Correa:2019rdk}. For example, the condition $\dot\alpha^m = 0$. This is like a Dirichlet boundary condition $\alpha^m = \alpha^m_0$ without specifying $\alpha^m_0$, and should be associated with the smeared configurations mentioned above. A different example of a delocalized string configuration is of course associated with Neumann boundary conditions.
This complementary analysis on supersymmetric boundary conditions for the fluctuations is presented in section \ref{section_fluctuations}. Many details concerning the derivation of the Killing spinors of the background, the action of quadratic fluctuations and their supersymmetries are relegated to the appendices \ref{killingspinors} and \ref{mass spectrum apendix}.

\section{Supersymmetric strings on $AdS_3\times S^3\times S^3\times S^1$}
\label{BPS classical strings}

We shall look for supersymmetric open strings ending  on the boundary of the type IIB supergravity background $AdS_3\times S^3\times S^3\times S^1$ with metric
\begin{equation}
 ds^2 =L^2 ds^2(AdS_3)+\frac{L^2}{\sin^2\Omega} ds^2(S^3_+)+ \frac{L^2}{\cos^2\Omega} ds^2(S^3_-)+l^2d\theta^2\,. 
\end{equation}
The parameter $\Omega$, that calibrates the radii of the 3-spheres, takes values in the range $0<\Omega<\frac{\pi}{2}$. This metric solves the type IIB supergravity equations of motion along with a constant dilaton $\Phi$ and a 3-form flux, which can be the Ramond-Ramond (R-R) one, the Neveu Schwarz-Neveu Schwarz (NS-NS) or a mixture of them. The R-R and NS-NS 3-form field strengths are
\begin{align}
F_{(3)} & = d C_{(2)} =  -2e^{-\Phi}L^2 \cos\lambda
\left({\rm vol}(AdS_3)+\frac{1}{\sin^2\Omega} {\rm vol}(S^3_+)+ \frac{1}{\cos^2\Omega} {\rm vol}(S^3_-) \right)\,,
\\
H_{(3)} & = d B_{(2)} = 2L^2 \sin\lambda
\left({\rm vol}(AdS_3)+\frac{1}{\sin^2\Omega} {\rm vol}(S^3_+)+ \frac{1}{\cos^2\Omega} {\rm vol}(S^3_-) \right)\,.
\end{align}
The parameter $\lambda$, taking values in the range $0\leq \lambda \leq \frac{\pi}{2}$, interpolates between the pure R-R and the pure NS-NS backgrounds. 

In the following we will parametrize $AdS_3$ with Poincar\'e coordinates
\begin{equation}
ds^2(AdS_3) =\frac{1}{z^2}\left(-dt^2+dx^2+dz^2\right)
    \qquad
    \Rightarrow
    \qquad {\rm vol}(AdS_3) =  \frac{1}{z^3}dt\wedge dx\wedge dz\,,
    \label{poincare}
\end{equation}
and the 3-spheres as
\begin{equation}
ds^2(S^3_\pm) =d\beta_\pm^2 + \cos^2 \beta_\pm \left(d\gamma_\pm+\cos^2\gamma_\pm d\varphi_\pm^2\right)
    \qquad
    \Rightarrow
    \qquad {\rm vol}(S_\pm^3) =  \cos^2\beta_\pm\cos\gamma_\pm d\beta_\pm\wedge d\gamma_\pm \wedge d\varphi_\pm\,.
    \label{spherecoor}
\end{equation}

The kind of open strings we shall consider  in first place are those ending at a straight line at the boundary of $AdS_3$ and sitting at fixed points in the compact spheres. An ansatz to describe them is the following:
\begin{equation}
t = \omega\tau\,,\qquad
x = x(\sigma)\,,\qquad 
z = \sigma\,,\qquad 
\beta_\pm,\gamma_\pm,\varphi_\pm,\theta = \text{const.}
\label{ansatz}
\end{equation}

To look for  supersymmetric string configurations we need the Killing spinors of the type IIB supergravity solutions we presented above. We give their detailed derivation in the appendix \ref{killingspinors}. Separating into real and imaginary parts $\epsilon = \eta + i\xi$ we find
\begin{align}
\eta & = U\epsilon_0 + \frac{\cos\lambda}{\sin\lambda + 1} V \epsilon_1
\label{etadef}
\\
\xi & = -\frac{\cos\lambda}{\sin\lambda + 1} U\epsilon_0 +  V \epsilon_1
\label{xidef}
\end{align}
where
\begin{align}
\label{U}
U & =  e^{\beta_+M_{\beta_+}}e^{\beta_-M_{\beta_-}}
e^{\gamma_+M_{\gamma_+}}e^{\gamma_-M_{\gamma_-}}e^{\varphi_+M_{\varphi_+}}e^{\varphi_-M_{\varphi_-}} e^{\log z M_z} e^{t (M_t+M_x)}e^{x (M_t+M_x)}\,,
\\
\label{V}
V & =  e^{-\beta_+M_{\beta_+}}e^{-\beta_-M_{\beta_-}}e^{-\gamma_+M_{\gamma_+}}e^{-\gamma_-M_{\gamma_-}}e^{-\varphi_+M_{\varphi_+}}e^{-\varphi_-M_{\varphi_-}}e^{-\log z M_z}e^{-t (M_t-M_x)}e^{x (M_t-M_x)}
\,,
\end{align}
which are defined in terms of the matrices
\begin{empheq}{alignat=9}
    M_t & = \tfrac{1}{2}\gamma^1\gamma^2 \,, &\qquad  M_x & = -\tfrac{1}{2}\gamma^0\gamma^2  \,, &\qquad & M_z & = \tfrac{1}{2}\gamma^0\gamma^1 \,, &
    \label{Mtxz}
    \\
    M_{\beta_+} & = \tfrac{1}{2}\gamma^4\gamma^5 \,, &\qquad  M_{\gamma_+} & =- \tfrac{1}{2}\gamma^3\gamma^5  \,, &\qquad & M_{\varphi_+} & = \tfrac{1}{2}\gamma^3\gamma^4  \,, &
\label{Mplus}
\\
M_{\beta_-} & = \tfrac{1}{2}\gamma^7\gamma^8 \,, &\qquad  M_{\gamma_-} & =- \tfrac{1}{2}\gamma^6\gamma^8  \,, &\qquad & M_{\varphi_-} & = \tfrac{1}{2}\gamma^6\gamma^7  \,, &
\label{Mminus}
    \end{empheq}

The constant spinors $\epsilon_0$ and $\epsilon_1$ are Majorana-Weyl spinors with the same chirality
\begin{equation}
\gamma_{11}\epsilon_0 = -\epsilon_0 \,,
\qquad
\gamma_{11}\epsilon_1 = -\epsilon_1 \,,
\end{equation}
and are further subject to an additional projection \eqref{killingcond1bis}
\begin{equation}
P^{\Sigma}_- \epsilon_0 = 0\,,\qquad
P^{\Sigma}_- \epsilon_1 = 0\,.
\end{equation}
This reduces the number of independent real parameters in the Killing spinors to 16.

A string configuration is supersymmetric if there are Killing spinors satisfying the $\kappa$-symmetry projection
\begin{equation}
\Gamma\epsilon = \epsilon\,,    
\label{kappapro0}
\end{equation}
where
\begin{equation}
    \Gamma = -\frac{\partial_\tau X^\mu \partial_\sigma X^\nu\Gamma_{\mu\nu}}{\sqrt{-h}}K\,.
\end{equation}
In this projector, $K$ stands for the complex conjugation operation and $h$ is the determinant of the induced metric. For the ansatz \eqref{ansatz} we obtain
\begin{equation}
\Gamma = -\frac{2}{\sqrt{1+(x')^2}}\left(M_x - x' M_z\right)K
:= \tilde \Gamma K\,.
\end{equation}

Separating into real and imaginary parts, the $\kappa$-symmetry projection \eqref{kappapro0} implies
\begin{align}
 (1-\tilde\Gamma) U\epsilon_0 & =  -\frac{\cos\lambda}{\sin\lambda+1}  (1-\tilde\Gamma) V\epsilon_1\,, 
 \\
 (1+\tilde\Gamma) U\epsilon_0 & =  \frac{\sin\lambda+1}{\cos\lambda}  (1+\tilde\Gamma) V\epsilon_1 \,,
\end{align}
and from their combination we obtain
\begin{equation}
\epsilon_0 = \tan\lambda\, U^{-1}V\epsilon_1 + \frac{1}{\cos\lambda} U^{-1}\tilde\Gamma V\epsilon_1\,.
\label{e0e1relation}
\end{equation}
Evaluated on the ansatz \eqref{ansatz}, the matrices defining the Killing spinors are
\begin{align}
U & =  U_7\, e^{\log \sigma M_z} e^{(x(\sigma)+\omega\tau)(M_t+M_x)}\,,
\\
V & =  V_7\, e^{-\log \sigma M_z} e^{(x(\sigma)-\omega\tau)(M_t-M_x)}\,,
\end{align}
where $U_7$ and $V_7$ are defined in the appendix. Both $U_7$ and $V_7$ commute with $M_t$, $M_x$ and $M_z$. Then, we get
\begin{align}
   U^{-1}V & = \frac{U_7^{-1}V_7}{2\sigma}\left[
   1-\omega^2\tau^2+x(\sigma)^2+\sigma^2
   -4 \omega\tau M_t-4x(\sigma) M_x +2(1+\omega^2\tau^2-x(\sigma)^2-\sigma^2)M_z
   \right]\,,
   \\
   U^{-1}\tilde\Gamma V & = 
  \frac{U_7^{-1}V_7}{2\sigma\sqrt{1+(x')^2}}\left[
   x'(1-\omega^2\tau^2+x(\sigma)^2-\sigma^2)+2\sigma x(\sigma) -4 x' \omega\tau M_t-
   4(x'x(\sigma) +\sigma)M_x\right. \nonumber
   \\
&  \hspace{2.6cm} \left. 
+(2x'(1+\omega^2\tau^2-x(\sigma)^2 +\sigma^2)-4\sigma x(\sigma))M_z\right]\,,
\end{align}
Implementing these expressions, equation \eqref{e0e1relation} becomes
\begin{align}
    \epsilon_0 & =
    \frac{U_7^{-1}V_7}{2\sigma\cos\lambda}
   \left[\tfrac{2\sigma x(\sigma)+x'\left(1+x(\sigma)^2 -\sigma^2\right)}{\sqrt{1+(x')^2}}+\left(1+x(\sigma)^2 +\sigma^2\right)\sin\lambda \right.
\nonumber\\
   & \hspace{1.7cm}
   \left.  -2\left(\tfrac{2\sigma x(\sigma)-x'\left(1-x(\sigma)^2 +\sigma^2\right)}{\sqrt{1+(x')^2}}-\left(1-x(\sigma)^2 -\sigma^2\right)\sin\lambda\right)M_z -4\left(
   \tfrac{ x'x(\sigma)+\sigma}{\sqrt{1+(x')^2}}+x(\sigma)\sin\lambda
   \right)M_x
   \right]\epsilon_1 \nonumber \\
   &\hspace{0.4cm}
   -\frac{U_7^{-1}V_7}{2\sigma\cos\lambda}
   \left[\tfrac{x'}{\sqrt{1+(x')^2}} +\sin\lambda\right] \left(4\omega\tau M_t+ (1-2M_z)\omega^2\tau^2\right)\epsilon_1\,. 
   \label{e0e1replaced}
     \end{align}

Since $\epsilon_0$ and $\epsilon_1$ are constant spinors, the string configuration will be supersymmetric if $x(\sigma)$ can be chosen so that all the dependence on $\sigma$ and $\tau$ goes away from the r.h.s. of \eqref{e0e1replaced}. Only the second term is $\tau$-dependent, which is cancelled provided that
\begin{equation}
    x'(\sigma) = -\tan\lambda\,,
    \qquad
    \Leftrightarrow
    \qquad 
    x(\sigma) = x_0 - \tan\lambda \ \sigma
    \,.
    \label{susyprofile}
\end{equation}
If this is the case, \eqref{e0e1replaced} becomes
\begin{align}
\label{susy constraint dirichlet}
    \epsilon_0 & =
    U_7^{-1}V_7\left(x_0 -2 x_0 M_z-2 M_x\right)\epsilon_1\,.
\end{align}

Therefore, the string configuration
\begin{equation}
t = \omega\tau\,,\qquad
x =  x_0 - \tan\lambda \ \sigma\,,\qquad 
z = \sigma\,,\qquad 
\beta_\pm,\gamma_\pm,\varphi_\pm,\theta = \text{const.}
\label{susyconfiguration}
\end{equation}
is 1/2 BPS. The worldsheet ends along the line $x = x_0$ at the boundary and it is tilt by an angle $-\lambda$. The induced geometry turns out to be that of and $AdS_2$ space. In what follows we will choose 
\begin{equation}
\label{omega}
\omega=\frac{1}{\cos \lambda}\,,
\end{equation}
so that the induced metric becomes
\begin{equation}
ds^2 = \frac{L^2 }{\sigma^2 \cos^2 \lambda }
\left(-d\tau^2 + d\sigma^2\right)\,,
\end{equation}
and the radius of the $AdS_2$ space is $R:=L/\cos \lambda$.
As one might have expected, the configuration \eqref{susyconfiguration} solves the equations of motion for an open string that couples to the NS-NS $B$-field (see Appendix \ref{mass spectrum apendix}). It is precisely this coupling the reason for the tilt. The configuration \eqref{susyconfiguration} has been previously found as a classical solution in \cite{Hernandez:2019huf}.

Finally, though in the rest of this paper we will focus on the case $0 \leq \lambda < \frac{\pi}{2}$, it is interesting to see what happens in the pure NS-NS limit. When $\lambda\to\frac\pi{2}$ the worldsheet becomes parallel to the boundary, so the limit of \eqref{susyconfiguration} is just one particular case of a more general solution of the type
\begin{equation}
\label{ns ns string}
t = \tau\,,\qquad
x = - \sigma\,,\qquad 
z = z_0 \,,\qquad 
\beta_\pm,\gamma_\pm,\varphi_\pm,\theta = \text{const.}
\end{equation}
In this case we get
\begin{equation}
\tilde{\Gamma}=-2 M_z\,,
\end{equation}
and the $\kappa$-symmetry projection \eqref{kappapro} translates into
\begin{align}
    \label{susy condition ns ns 1 0}
    2 U^{-1} M_z U \epsilon_0 &=-\epsilon_0\,,  \\
     \label{susy condition ns ns 2 0}
    2 V^{-1} M_z V \epsilon_1 &=\epsilon_1\,.
\end{align}
Using \eqref{U} and \eqref{V} we get
\begin{align}
    \label{susy condition ns ns 1}
    2 U^{-1} M_z U &= 2 M_z [1+2(M_t+M_x) (\tau-\sigma)]\,, \\
    \label{susy condition ns ns 2}
2 V^{-1} M_z V &= 2 M_z [1-2(M_t-M_x) (\tau+\sigma)]\,.
\end{align}
Thus, in order to remove the $\tau-$dependence in \eqref{susy condition ns ns 1} and \eqref{susy condition ns ns 2} and get a BPS string,  it suffices to impose
\begin{align}
    \label{susy constraint ns ns 1}
    (M_t+M_x) \; \epsilon_0 &=0\,, \qquad \Leftrightarrow \qquad 2 M_z \epsilon_0 = - \epsilon_0\,, \\
    \label{susy constraint ns ns 2}
   (M_t-M_x) \; \epsilon_1 &=0\,, \qquad \Leftrightarrow \qquad 2 M_z \epsilon_1 =  \epsilon_1\,.
\end{align}
The strings satisfying \eqref{ns ns string} preserve 8 independent Killing spinors for all values of $z_0$, and we get therefore a one-parameter family of 1/2 BPS strings which extend parallel to the boundary of $AdS_3$.
It is immediate to see that the induced metric for these strings is flat.

\subsection{Supersymmetric smeared strings}
\label{classical delocalized strings}

Generically, and because $U_7$ and $V_7$ depend on the values $\beta_\pm,\gamma_\pm,\varphi_\pm$,
different string configurations \eqref{susyconfiguration} at different points of the internal space $S^3\times S^3$ preserve different sets of supersymmetries. To conclude this section, we shall observe the existence of subspaces ${\cal M}\subset S^3\times S^3\times S^1$ such that all the strings sitting in ${\cal M}$ share some fraction of supersymmetry, which enables to have supersymmetric smeared string configurations. By smeared configurations we mean that the corresponding open string partition function should account for the superposition of strings with Dirichlet boundary conditions sitting at different points of a given ${\cal M}$.

One obvious possibility is to take ${\cal M}_1 = S^1$, as the Killing spinors are independent of the angle $\theta$. A  configuration of strings smeared along that circle should therefore be 1/2 BPS as well.This is in stark contrast with the analogue cases in $AdS_5\times S^5$ or $AdS_4\times CP^3$. Smearing open strings along compact subspaces in those cases breaks the supersymmetry either fully or partially.

Less obvious examples involve smearing in the $S^3\times S^3$ factor. The angular dependence of the projection \eqref{susy constraint dirichlet} is given by the matrix
\begin{equation}
\label{angular dependence matrix}
U_7^{-1} V_7= e^{-\varphi_+ M_{\varphi_+}-\varphi_- M_{\varphi_-}} e^{-\gamma_+ M_{\gamma_+}-\gamma_- M_{\gamma_-}} e^{-2 ( \beta_+ M_{\beta_+}+ \beta_- M_{\beta_-})} e^{-\gamma_+ M_{\gamma_+}-\gamma_- M_{\gamma_-}} e^{-\varphi_+ M_{\varphi_+}-\varphi_- M_{\varphi_-}}\,.
\end{equation}

As a concrete example, let us consider strings sitting in two maximal circles of the 3-spheres, by setting $\beta_\pm = \gamma_\pm = 0$. Then
\begin{equation}
\left. U_7^{-1} V_7\right|_{\beta_\pm = \gamma_\pm = 0} =  e^{-2 ( \varphi_+ M_{\varphi_+}+ \varphi_- M_{\varphi_-})}
= e^{-(\varphi_++\varphi_-)(M_{\varphi_+}+M_{\varphi_-})}
e^{-(\varphi_+-\varphi_-)(M_{\varphi_+}-M_{\varphi_-})}
\,,
\label{UV2circles}
\end{equation}
which implies that, for spinors also satisfying
\begin{equation}
    \label{delocalized string constraint 0}
    (M_{\varphi_+}+M_{\varphi_-}) \; \epsilon_1=0\,, \qquad \Leftrightarrow \qquad 4 M_{\varphi_+} M_{\varphi_-} \epsilon_1= \epsilon_1\,,
\end{equation}
the matrix \eqref{UV2circles} acts trivially on $\epsilon_1$ if the strings additionally satisfy that $\varphi_+-\varphi_-=0$. Thus, a second example is obtained by additionally smearing along this diagonal circle. Strings sitting at the submanifold ${\cal M}_2 = S^1_D\times S^1$
\begin{equation}
    \label{submanifold 1}
      \beta_{\pm} =  \gamma_{\pm} = 0\,,
      \qquad
     \varphi_+-\varphi_-=0\,,
\end{equation}
will be invariant under common supersymmetry transformations with 4 real parameters, because the additional projection \eqref{delocalized string constraint 0} halves the number of supersymmetries. More specifically,
\begin{align}
    \label{submanifold 1-non zero values-1}
     & (M_{\varphi_+}+M_{\varphi_-}) \; \epsilon_1 =0\,,\\
         \label{submanifold 1-non zero values-2}
     & \epsilon_0 =
     \left(x_0 -2 x_0 M_z-2 M_x\right) \; \epsilon_1\,.
\end{align}
In analogy with the case of type IIA strings on $AdS_4\times CP^3$ analyzed in \cite{Drukker:2008zx}, we conclude that a string smeared over the region $S^1_D\times S^1$ defined by \eqref{submanifold 1} is 1/4 BPS.
For different choices of the maximal circles one would encounter different projections but the smeared configurations will continue to be 1/4 BPS.

It is also possible to smear over larger submanifolds. A next example arises when we consider strings sitting in two maximal 2-spheres,  setting in this case $\beta_\pm = 0$. Smearing over ${\cal M}_3 = S^2_D\times S^1\supset {\cal M}_2$ , where the diagonal 2-sphere is defined by
\begin{equation}
    \label{submanifold 2}
      \beta_{\pm} =  0\,,
      \qquad
     \gamma_+-\gamma_-=0\,,
      \qquad
     \varphi_+-\varphi_-=0\,,
\end{equation}
the configuration is supersymmetric if, in addition to
\eqref{submanifold 1-non zero values-1} and \eqref{submanifold 1-non zero values-2}, one also imposes that
\begin{align}
    \label{submanifold 2-non zero values-1}
     & (M_{\gamma_+}+M_{\gamma_-}) \; \epsilon_1 =0\,.
\end{align}
With this additional projection the Killing spinors contain 2 real parameters, which means this smearing is 1/8 BPS.

As a final example, we can consider a string smeared over ${\cal M}_4 = S^3_D\times S^1\supset {\cal M}_3$ defined by
\begin{equation}
    \label{submanifold 3}
    \beta_+-\beta_-=0, \qquad
    \gamma_+-\gamma_-=0, \qquad 
    \varphi_+-\varphi_-=0. 
\end{equation}
To have common supersymmetries in this larger submanifold it should also be required that
\begin{align}
    \label{submanifold 3-non zero values-1}
    & (M_{\beta_+}+M_{\beta_-}) \; \epsilon_1 =0\,.
\end{align}
Notice, however, that 
\begin{equation}
\label{relation between delocalized string constraints}
4M_{\beta_+} M_{\beta_-} =(4 M_{\gamma_+} M_{\gamma_-} ) (4M_{\varphi_+} M_{\varphi_-})\,,
\end{equation}
which implies that  \eqref{submanifold 3-non zero values-1} is not an independent constraint with respect to \eqref{submanifold 1-non zero values-1} and \eqref{submanifold 2-non zero values-1}. Then, a string smeared over the larger ${\cal M}_4$ will also be 1/8 BPS.

\section{Fluctuations around classical solutions}
\label{section_fluctuations}

In this section, the fact that delocalized string configurations can be 1/2 BPS, 1/4 BPS or 1/8 BPS will be derived in an alternative way. More precisely, we will analyze the fluctuations around the classical solution \eqref{susyconfiguration}  up to quadratic order and study boundary conditions, other than Dirichlet, that can be consistent with supersymmetry. The mass spectrum for quantum corrections was calculated in \cite{Hernandez:2019huf} for the mixed-flux type IIB theory in $AdS_3\times S^3 \times T^4$. We should extend the analysis  to the $AdS_3\times S^3 \times S^3 \times S^1$ case. We do this in detail in the
Appendix \ref{mass spectrum apendix}. As shown there, a quadratic expansion of the action around \eqref{susyconfiguration} yields 7 massless scalars  corresponding to fluctuations in the directions of $S^{3}\times S^3 \times S^1$ and 1 massive scalar $\phi^{\sf tr}$, with mass $m_{\sf tr}^2=2/R^2$,  for the fluctuation transverse  to the worldsheet in $AdS_3$. The classical solution is sitting at a point in each of the 3-spheres, which breaks their isommetries $SO(4)$ down to $SO(3)$.  The fluctuations must therefore accommodate into representations of these residual symmetries. Three of the scalar fluctuations $\phi^{a}_{b}$ form a  $\square\!\square_+$ of the $SO(3)_+$\footnote{$a$,$b$ takes values 1 or 2 and $\phi^a_b$ is traceless.}, another three $\phi^{\dot{a}}_{\dot{b}}$ form a $\square\!\square_-$ of the $SO(3)_-$, and the remaining $\phi^{\sf tr}$ and $\phi^9$ are in the trivial representation of both $SO(3)$.

Concerning the fermionic fluctuations, after reducing them to $AdS_2$ spinors, we see that 4 of them are massless $\psi^{ a\dot{a}}$, while the remaining 4 $\chi^{a\dot{a}}$ have masses $m_{F}= 1/R$. These two sets transform in the
$\square_+\otimes\square_-$ of $SO(3)_+\times SO(3)_-$.

The action for the quadratic fluctuations can be then written as 
\begin{align}
\label{quadratic action expansion}
S= &-\frac{1}{2} \int d^2 \sigma \; \sqrt{-h} \left( \partial_{\alpha} \phi^{\sf tr} \partial^{\alpha} \phi^{\sf tr} + \frac{2}{R^2} (\phi^{\sf tr})^2 + 
\frac{1}{2}\partial_{\alpha} \phi^{a}_{b}
\partial^{\alpha} \phi^{b}_{a} 
+
\frac{1}{2}\partial_{\alpha} \phi^{\dot{a}}_{\dot{b}}
\partial^{\alpha} \phi^{\dot{b}}_{\dot{a}}+ \partial_{\alpha} \phi^{9} \partial^{\alpha} \phi^{9}
\right)\nonumber\\
&
-\frac{i}2 \int d^2 \sigma\; \sqrt{-h} \left(\bar{\psi}_{{a}\dot{a}} \ \slash\!\!\!\!{\cal D}^{(2)}  \psi^{{a}\dot{a}}  +
    \bar{\chi}_{{a}\dot{a}} \left( \slash\!\!\!\!{\cal D}^{(2)}
    -\frac{1}{R}\right)\chi^{{a}\dot{a}}\right)\,,
\end{align}
where $h_{\rm \alpha \beta}$ is the induced metric on the worldsheet, evaluated on the classical solution \eqref{susyconfiguration}, and $\mathcal{D}^{(2)}$ is the covariant derivative in $AdS_2$. In the following, the two-dimensional Dirac matrices will be denoted as $\uptau^0$ and $\uptau^1$, with $\uptau^3:=\uptau^0\uptau^1$ the chirality matrix.  The supersymmetry transformations of the action \eqref{quadratic action expansion}, derived in the appendix \ref{ads2 fields appendix},
 are
\begin{align}
\label{susy massive fermions-1}
\delta \chi^{a \dot{a}} &= \frac{1}{2} \left( \slashed{\partial} \phi^{\sf tr} + \frac{\phi^{\sf tr}}{R}  \right) \uptau^3 \kappa^{a\dot{a}} -\frac{i}{2} \left[ \sin \Omega \left( \slashed{\partial} \phi^{a}_{b} - \frac{\phi^{a}_{b}}{R} \right) \kappa^{b\dot{a}} + \cos \Omega \left( \slashed{\partial} \phi^{\dot{a}}_{\dot{b}} - \frac{\phi^{\rm \dot{a}}_{\dot{b}}}{R} \right) \kappa^{a\dot{b}} \right] \,, \\
\label{susy massless fermions-1}
\delta \psi^{a \dot{a}} &= \frac{1}{2} \slashed{\partial} \phi^{9} \kappa^{a \dot{a}}+\frac{i}{2} \left( \cos \Omega \; \slashed{\partial} \phi^{a}_{b} \kappa^{b \dot{a}}-\sin \Omega \; \slashed{\partial} \phi^{\dot{a}}_{\dot{b}} \kappa^{a \dot{b}} \right) \,, \\
\label{susy s1 scalar-1}
\delta \phi^{9} &=  -\frac{1}{2} \; \bar{ \psi}_{a \dot{a}}  \kappa^{a \dot{a}} \,, \\
\label{susy tr scalar-1}
\delta \phi^{\sf tr} &=  -\frac{1}{2} \; \bar{ \chi}_{\rm a \dot{a}} \uptau^3 \kappa^{a \dot{a}} \,, \\
\label{susy s3+ scalars-1}
\delta \phi^{a}_{b} &=-\frac{i}{2} \left[ \cos \Omega \left( 2 \; \bar{\psi}_{b\dot{c}} \kappa^{a\dot{c}} - \delta^{a}_{b} \bar{\psi}_{c\dot{c}} \kappa^{c\dot{c}} \right) -\sin \Omega \left(2 \; \bar{\chi}_{b\dot{c}} \kappa^{a\dot{c}} - \delta^{a}_{b} \bar{\chi}_{c\dot{c}} \kappa^{c\dot{c}} \right) \right] \,, \\
\label{susy s3- scalars-1}
\delta \phi^{\dot{a}}_{\dot{b}} &= \frac{i}{2} \left[ \sin \Omega \left( 2 \; \bar{\psi}_{c\dot{b}} \kappa^{c\dot{a}} - \delta^{\dot{a}}_{\dot{b}} \bar{\psi}_{c\dot{c}} \kappa^{c\dot{c}} \right) +\cos \Omega  \left(2 \; \bar{\chi}_{c\dot{b}} \kappa^{c\dot{a}} - \delta^{\dot{a}}_{\dot{b}} \bar{\chi}_{c\dot{c}} \kappa^{c\dot{c}} \right) \right] \,,
\end{align}
where $\kappa^{a \dot{a}}$ are $AdS_2$ Killing spinors, which satisfy
\begin{equation}
 \kappa^{a\dot{a}} =  \sigma^{-1/2} \varepsilon^{a\dot{a}} 
 +\sigma^{1/2} \uptau^0 \dot{\varepsilon}^{a\dot{a}} \,,
\end{equation}
with $\varepsilon(\tau)$ a spinor such that 
\begin{equation}
\uptau^1 \varepsilon^{{a}\dot{a}} =- \varepsilon^{{a}\dot{a}}\,, \qquad \quad \ddot{\varepsilon}^{{a}\dot{a}}=0 \,.
\end{equation}
These constraints, in addition to the property $\left( \varepsilon \right)^*_{a \dot{a}}= \epsilon_{a b} \epsilon_{\dot{a} \dot{b}} \; \varepsilon^{b \dot{b}}$ (where $\epsilon_{\dot{a} \dot{b}}$ and $\epsilon_{\dot{a} \dot{b}}$ are Levi-Civita tensors), imply that $\varepsilon^{a\dot a}$ parametrizes 8 real degrees of freedom, as expected.

\subsection{Supersymmetric boundary conditions}

In order to search for supersymmetric boundary conditions appropriate to describe BPS strings we shall now derive the asymptotic expansion of the supersymmetry transformations \eqref{susy massive fermions-1} to \eqref{susy s3- scalars-1}. Taking into account the usual $\sigma \to 0$ expansion for $AdS_2$ fields we get
\begin{align}
\label{chi expansion}
\chi^{a\dot{a}}(\tau,\sigma) &= \sigma^{-1/2} \left( \alpha_{\chi}^{a\dot{a}}(\tau) -\sigma \uptau^3 \dot{\alpha}_{\chi}^{a\dot{a}}(\tau) + \dots \right) +  \sigma^{3/2} \left( \beta_{\chi}^{a\dot{a}}(\tau) + \tfrac{1}{3}\sigma  \uptau^3\dot{\beta}_{\chi}^{a\dot{a}}(\tau) + \dots \right) \,, \\
\label{psi expansion}
\psi^{a\dot{a}} (\tau,\sigma) &= \sigma^{1/2} \left( \alpha_{\psi}^{a\dot{a}} (\tau) +\sigma \uptau^3 \dot{\alpha}_{\psi}^{a\dot{a}} (\tau) + \dots \right)+ \sigma^{1/2} \left( \beta_{\psi}^{a\dot{a}} (\tau) + \sigma \uptau^3 \dot{\beta}_{\psi}^{a\dot{a}} (\tau) + \dots \right) \,,  \\
\label{theta expansion}
\phi^{9} (\tau,\sigma) & = \left( \alpha_{9} (\tau)+ \dots \right) + \sigma \left( \beta_{9} (\tau) + \dots \right) \,, \\
\label{tr expansion}
\phi^{\sf tr} (\tau,\sigma) &= \sigma^{-1} \left( \alpha_{\sf tr} (\tau) + \dots \right) + \sigma^2 \left( \beta_{\sf tr} (\tau) + \dots \right) \,, \\
\label{phi s3+ expansion}
\phi^{a}_{b} (\tau,\sigma) &= \left( \alpha^{a}_{b} (\tau)+ \dots \right)+ \sigma \left( \beta^{a}_{b} (\tau) + \dots \right) \,, \\
\label{phi s3- expansion}
\phi^{\dot{a}}_{\dot{b}} (\tau,\sigma) &= \left( \alpha^{\dot{a}}_{\dot{b}} (\tau)+ \dots \right) + \sigma \left( \beta^{\dot{a}}_{\dot{b}} (\tau) + \dots \right) \,,
\end{align}
for the fermionic and bosonic fluctuations, where
\begin{align}
\uptau^{1} \alpha_{\chi}^{a\dot{a}}=- \alpha_{\chi}^{a\dot{a}} \,, \qquad \uptau^{1} \beta_{\chi}^{a\dot{a}}= \beta_{\chi}^{a\dot{a}} \,, \\
\uptau^{1} \alpha_{\psi}^{a\dot{a}}=- \alpha_{\psi}^{a\dot{a}}\,,  \qquad \uptau^{1} \beta_{\psi}^{a\dot{a}}= \beta_{\psi}^{a\dot{a}} \,.
\end{align}

Due to the Breitenlohner-Freedman bound \cite{Breitenlohner:1982jf,Breitenlohner:1982bm} we must always take
\begin{align}
\label{bf bc 1}
\alpha_{\chi}^{a\dot{a}}=0 \quad \forall a,\dot{a} \,; 
\qquad
\alpha_{\sf tr}=0 \,,
\end{align}
as boundary conditions for the $\chi^{a\dot{a}}$ and $\phi^{\sf tr}$ fields. However, we have some freedom to impose more general boundary conditions on the other fluctuations. In the following, conditions such as $\alpha_{\chi}^{a\dot{a}}=0 \; \, \forall a,\dot{a}$ will be written simply as $\alpha_{\chi}^{a\dot{a}}=0$, without explicitly specifying that indices without contraction take all possible values.

Then, taking \eqref{bf bc 1} into consideration  
and inserting \eqref{chi expansion}-\eqref{phi s3- expansion} into the transformations \eqref{susy massive fermions-1}-\eqref{susy s3- scalars-1}, the transformations relevant to the analysis of supersymmetric boundary conditions become

\begin{align}
\label{susy alpha chi}
\delta \alpha_{\chi}^{a\dot{a}} &= \frac{i}{2R} \left( \sin \Omega \; \alpha^a_b \varepsilon^{b\dot{a}}+ \cos \Omega \; \alpha^{\dot{a}}_{\dot{b}} \varepsilon^{a\dot{b}} \right) \hspace{8cm}\\
\label{susy alpha psi}
\delta \alpha_{\psi}^{a\dot{a}} &= - \frac{1}{2R} \beta_{9} \varepsilon^{a\dot{a}}  - \frac{i}{2R} \left( \cos \Omega \; \beta^{a}_b \varepsilon^{b\dot{a}} - \sin \Omega \; \beta^{\dot{a}}_{\dot{b}} \varepsilon^{a\dot{b}} \right)  \\
\label{susy beta psi}
\delta \beta_{\psi}^{a\dot{a}} &= - \frac{\uptau^3}{2R} \dot{\alpha}_{9} \varepsilon^{a\dot{a}} - \frac{i}{2R} \uptau^3 \left( \cos \Omega \; \dot{\alpha}^{a}_b \varepsilon^{b\dot{a}} - \sin \Omega \; \dot{\alpha}^{\dot{a}}_{\dot{b}} \varepsilon^{a\dot{b}} \right) \\
\label{susy alpha 9}
\delta \alpha_{9} & = -\frac{1}{2} \overline{\beta}_{\psi,a\dot{a}} \varepsilon^{a\dot{a}}\\
\label{susy beta 9}
\delta \beta_{9} & = \frac{1}{2} \left( \dot{\overline{\alpha}}_{\psi,a\dot{a}} \uptau^3 \varepsilon^{a\dot{a}} + \overline{\alpha}_{\psi,a\dot{a}} \uptau^3 \dot{\varepsilon}^{a\dot{a}} \right) \\
\label{susy alpha tr}
\delta \alpha_{\sf tr} &= 0 \\
\label{susy alpha su2+}
\delta \alpha^a_b &= -\frac{i \cos \Omega}{2} \left( 2 \overline{\beta}_{\psi,b\dot{c}} \varepsilon^{a\dot{c}} -\delta^a_b \overline{\beta}_{\psi,c\dot{c}} \varepsilon^{c\dot{c}} \right) \\
\label{susy beta su2+}
\delta \beta^a_b &= -\frac{i \cos \Omega}{2} \left[ -2 \left(  \dot{\overline{\alpha}}_{\psi,b\dot{c}} \uptau^3 \varepsilon^{a\dot{c}} + \overline{\alpha}_{\psi,b\dot{c}} \uptau^3 \dot{\varepsilon}^{a\dot{c}} \right) +\delta^a_b \left(  \dot{\overline{\alpha}}_{\psi,c\dot{c}} \uptau^3 \varepsilon^{c\dot{c}} + \overline{\alpha}_{\psi,c\dot{c}} \uptau^3 \dot{\varepsilon}^{c\dot{c}} \right) \right] \nonumber \\
& \hspace{0.40cm} +\frac{i \sin \Omega}{2} \left( 2 \overline{\beta}_{\chi,b\dot{c}} \varepsilon^{a\dot{c}}-\delta^a_b \overline{\beta}_{\chi,c\dot{c}} \varepsilon^{c\dot{c}} \right) \\
\label{susy alpha su2-}
\delta \alpha^{\dot{a}}_{\dot{b}} &= \frac{i \sin \Omega}{2} \left( 2 \overline{\beta}_{\psi,c\dot{b}} \varepsilon^{c\dot{a}} -\delta^{\dot{a}}_{\dot{b}} \overline{\beta}_{\psi,c\dot{c}} \varepsilon^{c\dot{c}} \right) \\
\label{susy beta su2-}
\delta \beta^{\dot{a}}_{\dot{b}} &=  \frac{i \sin \Omega}{2} \left[ -2 \left(  \dot{\overline{\alpha}}_{\psi,c\dot{b}} \uptau^3 \varepsilon^{c\dot{a}} + \overline{\alpha}_{\psi,c\dot{b}} \uptau^3 \dot{\varepsilon}^{c\dot{a}} \right) +\delta^{\dot{a}}_{\dot{b}} \left(  \dot{\overline{\alpha}}_{\psi,c\dot{c}} \uptau^3 \varepsilon^{c\dot{c}} + \overline{\alpha}_{\psi,c\dot{c}} \uptau^3 \dot{\varepsilon}^{c\dot{c}} \right) \right] \nonumber \\
& \hspace{0.40cm}  + \frac{i \cos \Omega}{2} \left( 2 \overline{\beta}_{\chi,c\dot{b}} \varepsilon^{c\dot{a}}-\delta^{\dot{a}}_{\dot{b}} \overline{\beta}_{\chi,c\dot{c}} \varepsilon^{c\dot{c}} \right)
\end{align}

\subsubsection{Dirichlet boundary conditions}

We will say that a set of boundary conditions is supersymmetric when the variation of them under \eqref{susy alpha chi}-\eqref{susy beta su2-} is vanishing for some  non-trivial choice of parameters $\varepsilon^{a\dot a}$.
It is straightforward to see from these expressions that Dirichlet boundary conditions
\begin{equation}
\alpha_{\chi}^{a\dot{a}}=  0\,, \qquad
\beta_{\psi}^{a\dot{a}} =  0\,, \qquad 
{\alpha}_{9} = 0  \,,\qquad
\alpha^{\dot{a}}_{\dot{b}} =  0\,, \qquad 
\alpha^a_b =  0\,,  \qquad  \alpha_{\sf tr} = 0 \,,
\end{equation}
preserve the full set of supersymmetries of the action (\ref{quadratic action expansion}), and thus matches the fact that the string given by (\ref{susyconfiguration}) is 1/2 BPS\footnote{All the supersymmetries of the action for quadratic fluctuations are one half of the supersymmetries of the background.}.  

\subsubsection{Smeared boundary conditions}
\label{smeared bc}

Let us now discuss the possibility of having delocalized supersymmetric string configurations. We presented various instances of supersymmetric smeared strings in section \ref{BPS classical strings}. It should also be possible to describe all of them in terms boundary conditions for the fluctuations. 

Let us first focus on the case of strings delocalized over the submanifold $\mathcal{M}_1=S^1$. As discussed above, the partition function of a string smeared over a submanifold $\mathcal{M}$ can be obtained as an integral over the partition functions of Dirichlet strings sitting on $\mathcal{M}$. In terms of the fluctuations, a \emph{uniform} smearing over $\mathcal{M}_1$ should correspond to imposing $\alpha_9 = \text{const.}$ and integrating over the constant, which is equivalent to imposing $\dot\alpha_9 = 0$.
Therefore, our proposal is that the string uniformly smeared over ${\cal M}_1$ should be described by the following boundary conditions on the fluctuations:
\begin{equation}
\label{smeared bc s1}
\alpha_{\chi}^{a\dot{a}}=  0\,, \qquad
\beta_{\psi}^{a\dot{a}} =  0\,, \qquad 
\dot{\alpha}_{9} = 0  \,,\qquad
\alpha^{\dot{a}}_{\dot{b}} =  0\,, \qquad 
\alpha^a_b =  0\,,  \qquad  \alpha_{\sf tr} = 0 \,.
\end{equation}
By looking at the transformations \eqref{susy alpha chi}-\eqref{susy beta su2-} we see that these boundary conditions are invariant under the action of the 8 supercharges parametrized by the $\varepsilon^{a \dot{a}}$
spinors, in agreement with the results obtained in section \ref{BPS classical strings} when analyzing the classical limit of this delocalized string.

On the other hand, for the string smeared over the submanifold $\mathcal{M}_2 = S^1_D\times S^1$ given at \eqref{submanifold 1} we must impose now
\begin{equation}
\label{smeared bc 1-zeta fluctuations}
\zeta^{\beta_{\pm}}=0 \,, \qquad
\zeta^{\gamma_{\pm}}=0 \,, \qquad
\zeta^{\varphi_+}-\zeta^{\varphi_-}=0 \,, \qquad
\dot{\zeta}^{\theta}=0 \,,
\end{equation}
on the $\zeta^{\mu}$ fluctuations defined in Appendix \ref{mass spectrum apendix}. Using the $\phi^a_b$ and $\phi^{\dot{a}}_{\dot{b}}$ fields (see \eqref{definition phi su2+} and \eqref{definition phi su2-} for a definition in terms of $\phi^m=e^m_{\mu} \zeta^{\mu}$) these boundary conditions can be expressed as
\begin{equation}
\label{smeared bc 1-scalar fluctuations}
\alpha^{1}_{2}=\alpha^{2}_{1}=0 \,, \qquad 
\alpha^{\dot{1}}_{\dot{2}}=\alpha^{\dot{2}}_{\dot{1}}=0 \,, \qquad 
\sin \Omega \; \alpha^{1}_{1}-\cos \Omega \; \alpha^{\dot{1}}_{\dot{1}}=0 \,, \qquad
\dot{\alpha}_9=0 \,,
\end{equation}
where the $\sin \Omega$ and $\cos \Omega$ factors come from the vielbein that relates the $\zeta^{\mu}$ fields with the scalar fluctuations (we have assumed the classical string to be sitting at the origin). Conditions \eqref{smeared bc 1-scalar fluctuations} have to be supplemented with a condition that accounts for the smearing along the diagonal $S^1_D$. This condition is simply written as 
$\cos \Omega \; \dot{\alpha}^{1}_{1}+\sin \Omega \; \dot{\alpha}^{\dot{1}}_{\dot{1}}=0$.

Looking again at the transformations \eqref{susy alpha chi}-\eqref{susy beta su2-}, we see that 
\begin{equation}
\label{smeared bc 1-complete}
\begin{aligned}
 \alpha_{\chi}^{a\dot{a}} &=0\,, \quad \; \,
 & \beta_{\psi}^{a\dot{a}}&=0\,, \quad & \alpha_{\sf tr}&=0\,,   \qquad & \dot{\alpha}_{9}&=0 \,, \\
\alpha^{1}_{2}=\alpha^{2}_{1} &=0 \,,  \qquad  &
\alpha^{\dot{1}}_{\dot{2}}=\alpha^{\dot{2}}_{\dot{1}}&=0 \,, \qquad &
\sin \Omega \; \alpha^{1}_{1}-\cos \Omega \; \alpha^{\dot{1}}_{\dot{1}}&=0 \,, \qquad &
\cos \Omega \; \dot{\alpha}^{1}_{1}+\sin \Omega \; \dot{\alpha}^{\dot{1}}_{\dot{1}}&=0 \,,
\end{aligned}
\end{equation}
define a  set of boundary conditions, which are invariant only under the supersymmetries parametrized by the spinors $\varepsilon^{1 \dot{2}}$ and $\varepsilon^{2 \dot{1}}$. The latter becomes clear when demanding that $\delta \alpha_{\chi}^{a \dot{a}}$ must be 0. 
Thus, the boundary conditions \eqref{smeared bc 1-complete} preserve 4 real supercharges, in agreement with the results presented in section \ref{BPS classical strings} for the 1/4 BPS classical string smeared over $\mathcal{M}_2$.

As for the string smeared over the submanifold $\mathcal{M}_3 = S^2_D\times S^1$ given at \eqref{submanifold 2}, in this case we must impose
\begin{equation}
\label{smeared bc 2-zeta fluctuations}
\begin{aligned}
& \zeta^{\beta_{\pm}}=0 \,, \qquad
\zeta^{\gamma_+}-\zeta^{\gamma_-}=0 \,, \qquad
\zeta^{\varphi_+}-\zeta^{\varphi_-}=0 \,, \qquad
\dot{\zeta}^{\theta}=0 \,,
\end{aligned}
\end{equation}
or, equivalently,
\begin{equation}
\label{smeared bc 2-scalar fluctuations}
\begin{aligned}
\alpha^{1}_{2}+ \alpha^{2}_{1}=0 \,, \qquad 
\alpha^{\dot{1}}_{\dot{2}}+ \alpha^{\dot{2}}_{\dot{1}}=0 \,, \quad
\sin \Omega \; \alpha^{1}_{2}-\cos \Omega \; \alpha^{\dot{1}}_{\dot{2}}=0 \,, \quad
\sin \Omega \; \alpha^{1}_{1}-\cos \Omega \; \alpha^{\dot{1}}_{\dot{1}}=0 \,, \quad 
\dot{\alpha}_9=0 \,.
\end{aligned}
\end{equation}
The conditions associated to the smearing are in this case $\cos \Omega \; \dot{\alpha}^{1}_{2}+\sin \Omega \; \dot{\alpha}^{\dot{1}}_{\dot{2}}=0$ and $\cos \Omega \; \dot{\alpha}^{1}_{1}+\sin \Omega \; \dot{\alpha}^{\dot{1}}_{\dot{1}}=0$. Then, the complete set of boundary conditions in this case is
\begin{equation}
\label{smeared bc 2-complete}
\begin{aligned}
\alpha_{\chi}^{a\dot{a}} &=0\,, \quad \; \,
 & \beta_{\psi}^{a\dot{a}}&=0\,, \quad & \alpha_{\sf tr}=0\,,   \qquad  \dot{\alpha}_{9}&=0 \,, 
 \\
\alpha^{1}_{2}+ \alpha^{2}_{1} &=0 \,, \qquad &
\sin \Omega \; \alpha^{1}_{2}-\cos \Omega \; \alpha^{\dot{1}}_{\dot{2}} &=0 \,, \qquad &
\cos \Omega \; \dot{\alpha}^{1}_{2}+\sin \Omega \; \dot{\alpha}^{\dot{1}}_{\dot{2}} &=0 \,, \\
\alpha^{\dot{1}}_{\dot{2}}+ \alpha^{\dot{2}}_{\dot{1}} &=0 \,,&
\sin \Omega \; \alpha^{1}_{1}-\cos \Omega \; \alpha^{\dot{1}}_{\dot{1}}&=0 \,, \quad &
\cos \Omega \; \dot{\alpha}^{1}_{1}+\sin \Omega \; \dot{\alpha}^{\dot{1}}_{\dot{1}}&=0 \,,
\end{aligned}
\end{equation}
and they are invariant under \eqref{susy alpha chi}-\eqref{susy beta su2-} for transformations
 which satisfy
\begin{equation}
\label{constraints susy 1/8 BPS}
\varepsilon^{1 \dot{1}}=\varepsilon^{2 \dot{2}}=0 \,, \qquad \varepsilon^{1 \dot{2}}=-\varepsilon^{2 \dot{1}} \;. 
\end{equation}
These constraints are solved by 2 real supercharges, and therefore the conditions \eqref{smeared bc 2-complete} match the results obtained previously for the classical string smeared over $\mathcal{M}_3$.

The boundary conditions \eqref{smeared bc 2-complete} can be slightly modified in order to describe the string smeared over the submanifold $\mathcal{M}_4 = S^3_D\times S^1$. In this case it suffices to impose
\begin{equation}
\label{smeared bc 3-complete}
\begin{aligned}
\alpha_{\chi}^{a\dot{a}} =0\,, \qquad \beta_{\psi}^{a\dot{a}}&=0\,, \quad & \alpha_{\sf tr}=0\,,   \qquad  \dot{\alpha}_{9}&=0 \,, 
\\
\sin \Omega \; \alpha^{1}_{2}-\cos \Omega \; \alpha^{\dot{1}}_{\dot{2}} &=0 \,, \qquad &
\cos \Omega \; \dot{\alpha}^{1}_{2}+\sin \Omega \; \dot{\alpha}^{\dot{1}}_{\dot{2}} &=0 \,, \\
\sin \Omega \; \alpha^{1}_{1}-\cos \Omega \; \alpha^{\dot{1}}_{\dot{1}}&=0 \,, \quad &
\cos \Omega \; \dot{\alpha}^{1}_{1}+\sin \Omega \; \dot{\alpha}^{\dot{1}}_{\dot{1}}&=0 \,,
\end{aligned}
\end{equation}
These conditions are again preserved only by the supersymmetries which solve \eqref{constraints susy 1/8 BPS}, in agreement with the results expected for the 1/8 BPS string smeared over $\mathcal{M}_4$.

\subsubsection{Neumann boundary conditions}
\label{neumann bc}

The analysis of fluctuations not only appears to be consistent with the results obtained when studying classical strings, but also enables us to characterize some other delocalized supersymmetric strings not accounted in section \ref{classical delocalized strings}. Instead of smearing over regions ${\cal M}\subset S^3\times S^3\times S^1$ we can also impose Neumann boundary conditions on the coordinates spanning them and ask in what cases they can be supersymmetric.

We can start by considering the imposition of a Neumann boundary condition for the fluctuation along ${\cal M}_1$ and Dirichlet for the rest
\begin{equation}
\label{neumann bc s1}
  \beta_{9} = 0  \,,
   \qquad  \alpha^a_b =  0\,,  \qquad  \alpha^{\dot{a}}_{\dot{b}}=  0\,, 
 \qquad  \alpha_{\sf tr} = 0 \,.
\end{equation}
A quick inspection of the transformations \eqref{susy alpha chi} to \eqref{susy beta su2-} reveals that 
\eqref{neumann bc s1} cannot preserve all the supersymmetries. In order to have $\delta\beta_9 = 0$ with the most general parameters the vanishing of all  $\alpha_\psi^{a\dot{a}}$ would be needed. 
The variation of the latter depend on $\beta^a_b$ and
$\beta^{\dot a}_{\dot b}$, which are not necessarily vanishing when imposing Dirichlet boundary conditions on $\phi^a_b$ and $\phi^{\dot a}_{\dot b}$. 

Nonetheless, \eqref{neumann bc s1} can preserve some fraction of the supersymmetry if for the fermionic fluctuations we demand that
\begin{equation}
\label{neumann bc s1fer}
\alpha_{\chi}^{a\dot{a}} = 0\,,\qquad
\alpha_{\psi}^{1\dot{2}}-\alpha_{\psi}^{2\dot{1}}=0\,,
\qquad
\beta_{\psi}^{1\dot{2}}+\beta_{\psi}^{2\dot{1}}=0\,,
\qquad 
\beta_{\psi}^{1\dot{1}}=\beta_{\psi}^{2\dot{2}}=0\,.
\end{equation}
In particular, \eqref{neumann bc s1} and \eqref{neumann bc s1fer} are invariant under transformations with $\varepsilon^{1 \dot{1}}=\varepsilon^{2 \dot{2}}=0$  and $ \varepsilon^{1 \dot{2}}=-\varepsilon^{2 \dot{1}}$. This example represents a notorious difference between the $AdS_3 \times S^3 \times S^3 \times S^1$ and the $AdS_4 \times CP^3$ cases. In this case, imposing $\dot{\alpha}=0$ or $\beta=0$ on the fluctuations of a given submanifold does not preserve the same amount of supersymmetry. This reinforces our proposal that the delocalized configurations presented in section \ref{classical delocalized strings} are accounted in terms of boundary conditions that fix $\dot{\alpha}$ for the corresponding scalar fluctuations.

Imposing Neumann boundary conditions for the fluctuation along ${\cal M}_2$ can be supersymmetric if some fermionic boundary conditions are accordingly changed. More precisely, from \eqref{susy alpha chi} to \eqref{susy beta su2-} we find that the boundary conditions
\begin{equation}
\label{extra bc 1-complete}
\begin{aligned}
  \alpha_{\sf tr}&=0\,,   \qquad & \beta_{9}&=0 \,,
  & \alpha_{\chi}^{a\dot{a}} &=0\,, 
 &\beta_{\psi}^{1\dot{1}}=\beta_{\psi}^{2\dot{2}}=\alpha_{\psi}^{1\dot{2}}=\alpha_{\psi}^{2\dot{1}}&=0\,, 
 \\
\alpha^{1}_{2}=\alpha^{2}_{1} &=0 \,,  \qquad  &
\alpha^{\dot{1}}_{\dot{2}}=\alpha^{\dot{2}}_{\dot{1}}&=0 \,, \qquad &
\sin \Omega \; \alpha^{1}_{1}-\cos \Omega \; \alpha^{\dot{1}}_{\dot{1}}&=0 \,, \qquad &
\cos \Omega \; \beta^{1}_{1}+\sin \Omega \; \beta^{\dot{1}}_{\dot{1}}&=0 \,,
\end{aligned}
\end{equation}
are invariant under the supersymmetries parametrized by $\varepsilon^{1\dot{2}}$ and $\varepsilon^{2\dot{1}}$, and thus describe a 1/4 BPS string. In contrast to the previous example, in this case the supersymmetries preserved are exactly the same as 
the ones preserved in the smearing over ${\cal M}_2$.

When turning to Neumann boundary conditions in ${\cal M}_3$, we can impose
\begin{equation}
\label{extra bc M3}
\begin{aligned}
\alpha_{\chi}^{a\dot{a}}&=0 \,, \qquad &\alpha_{\psi}^{1\dot{2}}=\alpha_{\psi}^{2\dot{1}}&=0 \,,\\
\alpha_{\psi}^{1\dot{1}}+\alpha_{\psi}^{2\dot{2}}&=0 \,, \qquad &\beta_{\psi}^{1\dot{1}}-\beta_{\psi}^{2\dot{2}}&=0 \,,\\
\alpha^1_2+\alpha^2_1=\alpha^{\dot{1}}_{\dot{2}}+\alpha^{\dot{2}}_{\dot{1}}&=0 \,, \qquad
&\alpha_{\sf tr}=0 \,, \quad \beta_9&=0 \,,\\
\sin \Omega  \left( \alpha^{1}_{2}- \alpha^{2}_{1} \right) - \cos \Omega \left( \alpha^{\dot{1}}_{\dot{2}} - \alpha^{\dot{2}}_{\dot{1}} \right) &=0 \,, \qquad
&\sin \Omega \; \alpha^{1}_{1} - \cos \Omega \; \alpha^{\dot{1}}_{\dot{1}}&=0 \,,\\
\cos \Omega  \left( \beta^{1}_{2}- \beta^{2}_{1} \right) + \sin \Omega \left( \beta^{\dot{1}}_{\dot{2}} - \beta^{\dot{2}}_{\dot{1}} \right) &=0 \,, \qquad
&\cos \Omega \; \beta^{1}_{1} + \sin \Omega \; \beta^{\dot{1}}_{\dot{1}}&=0 \,.
\end{aligned}
\end{equation}
These conditions preserve the supersymmetries which satisfy \eqref{constraints susy 1/8 BPS} and therefore describe a 1/8 BPS string, as in the case \eqref{smeared bc 3-complete} for the smearing over $\mathcal{M}_3$.

Our final example is the case of Neumann boundary conditions for the fluctuations along ${\cal M}_4$. In that case, the conditions
\begin{equation}
\label{extra bc 2-complete}
\begin{aligned}
\alpha_{\chi}^{a\dot{a}} =0\,, \quad \; \,
 \alpha_{\psi}^{a\dot{a}}&=0\,, \quad & \alpha_{\sf tr}=0\,,   \qquad  \beta_{9}&=0 \,, 
\\
\sin \Omega \; \alpha^{1}_{2}-\cos \Omega \; \alpha^{\dot{1}}_{\dot{2}} &=0\,, \qquad &
\sin \Omega \; \alpha^{1}_{1}-\cos \Omega \; \alpha^{\dot{1}}_{\dot{1}} &=0\,, \qquad &
\\
\cos \Omega \; \beta^{1}_{2}+\sin \Omega \; \beta^{\dot{1}}_{\dot{2}} &=0 \,, &
\cos \Omega \; \beta^{1}_{1}+\sin \Omega \; \beta^{\dot{1}}_{\dot{1}} &=0 \,, &
\end{aligned}
\end{equation}
preserve only the supersymmetries which satisfy \eqref{constraints susy 1/8 BPS}, and thus correspond to a 1/8 BPS string.

\subsubsection{Interpolations}

In the previous sections we have found diverse sets of 1/2 BPS, 1/4 BPS and 1/8 BPS strings which are described either by Dirichlet, smeared or Neumann boundary conditions. As we shall see in the following, there is a network of supersymmetric boundary conditions that allows us to interpolate between all those BPS strings.

Let us begin our analysis by studying the interpolation between the Dirichlet and smeared strings discussed in the previous sections. For this it is useful to consider first the case of the string smeared over $\mathcal{M}_1=S^1$, and to note that a Dirichlet string can be thought as a limiting case of such delocalized configuration. For the uniformly smeared string the partition function can be obtained by taking Dirichlet strings with boundary condition $\phi^9=\text{const.}$ and then integrating over the values of such constant.  We could in principle perform this integration with an arbitrary weight function and the configuration would continue to be supersymmetric. We can then think of  the Dirichlet string as the limit in which the region of support of the weight function collapses to a point. Thus, by smoothly deforming the domain of support we can interpolate between the Dirichlet string and the string uniformly smeared over $\mathcal{M}_1$. This idea can be generalized to interpolate between any of the strings considered in section \ref{BPS classical strings}. These configurations preserve the same amount of supersymmetry as the least supersymmetric endpoint of the interpolation and describe a marginal deformation
in the dual dCFT.

We can also interpolate between the smeared strings of section \ref{smeared bc} and the Neumann strings analyzed at section \ref{neumann bc}. This is similar to the mixed boundary conditions presented in \cite{Correa:2019rdk} for the $AdS_4 \times CP^3$ case. For example, we can interpolate between the smeared and Neumann boundary conditions over $\mathcal{M}_1$ if we impose
\begin{equation}
\label{interpolating bc M1}
\begin{aligned}
\alpha_{\chi}^{a\dot{a}}&=  0\,, \qquad &\beta_{\psi}^{1\dot{1}}=\beta_{\psi}^{2\dot{2}}&=0\,, \\
\beta_{\psi}^{1\dot{2}}+\beta_{\psi}^{2\dot{1}}&=0\,, \qquad &\left( \Lambda \, \alpha_{\psi}^{1\dot{2}} +\uptau^3 \beta_{\psi}^{1\dot{2}} \right)-\left( \Lambda \, \alpha_{\psi}^{2\dot{1}} +\uptau^3 \beta_{\psi}^{2\dot{1}} \right)&=0\,, \\
\alpha^a_b =\alpha^{\dot{a}}_{\dot{b}} =  0\,, \quad \alpha_{\sf tr} &= 0 \,, \qquad &\Lambda \, \beta_{9} + \dot{\alpha}_{9} &= 0  \,,
\end{aligned}
\end{equation}
where $\Lambda \in \mathbb{R}$. As expected, this set of conditions is invariant under 2 of the supersymmetries of the action \eqref{quadratic action expansion} for $\Lambda>0$, and in the limit $\Lambda=0$ preserves all the supersymmetries. Moreover, $\Lambda$ is a dimensionless parameter and, from the perspective of the 1$d$ dual dCFT, this interpolating boundary condition describes a superconformal marginal deformation.

In a similar way we can interpolate between all the smeared and Neumann boundary conditions presented in the previous sections. All those interpolating conditions preserve the same amount of supersymmetry as the least supersymmetric end of the flow, and describe marginal deformations in the dual dCFT.

\begin{figure}[h!]
\begin{center}
\begin{tikzpicture}
\node [BPSbox] at (1.05,0) {\small Dirichlet string \\ {\bf 1/2 BPS}};
\node [BPSbox2] at (3.5,1.2) {\small Smearing over ${\cal M}_1$ \\ {\bf 1/2 BPS}};
\node [BPSbox2] at (7.3,1.2) {\small Smearing over ${\cal M}_2$ \\ {\bf 1/4 BPS}};
\node [BPSbox2] at (11.1,1.2) {\small Smearing over ${\cal M}_3$ \\ {\bf 1/8 BPS}};
\node [BPSbox2] at (14.9,1.2) {\small Smearing over ${\cal M}_4$ \\ {\bf 1/8 BPS}};
\node [BPSbox2] at (3.5,-1.2) {\small Neumann on ${\cal M}_1$ \\ {\bf 1/8 BPS}};
\node [BPSbox2] at (7.3,-1.2) {\small Neumann on ${\cal M}_2$ \\ {\bf 1/4 BPS}};
\node [BPSbox2] at (11.1,-1.2) {\small Neumann on ${\cal M}_3$ \\ {\bf 1/8 BPS}};
\node [BPSbox2] at (14.9,-1.2) {\small Neumann on ${\cal M}_4$ \\ {\bf 1/8 BPS}};
\draw[<->,thick] (5,1.2)--(5.8,1.2);
\draw[<->,thick] (8.8,1.2)--(9.6,1.2);
\draw[<->,thick] (12.6,1.2)--(13.4,1.2);
\draw[<->,thick] (5,-1.2)--(5.8,-1.2);
\draw[<->,thick] (8.8,-1.2)--(9.6,-1.2);
\draw[<->,thick] (12.6,-1.2)--(13.4,-1.2);
\draw[<->,thick] (0.95,0.65)--(2,1.15);
\draw[<->,thick] (0.95,-0.65)--(2,-1.15);
\draw[<->,thick] (3.5,0.55)--(3.5,-0.55);
\draw[<->,thick] (7.3,0.55)--(7.3,-0.55);
\draw[<->,thick] (11.1,0.55)--(11.1,-0.55);
\draw[<->,thick] (14.9,0.55)--(14.9,-0.55);
\end{tikzpicture}
\caption{Examples of BPS delocalized string configurations. Arrows indicate BPS interpolations between them. The submanifolds are ${\cal M}_1 = S^1$, ${\cal M}_2 = S^1_D\times S^1$, ${\cal M}_3 = S^2_D\times S^1$ and ${\cal M}_4 = S^3_D\times S^1$.}
\label{resumefig}
\end{center}
\end{figure}
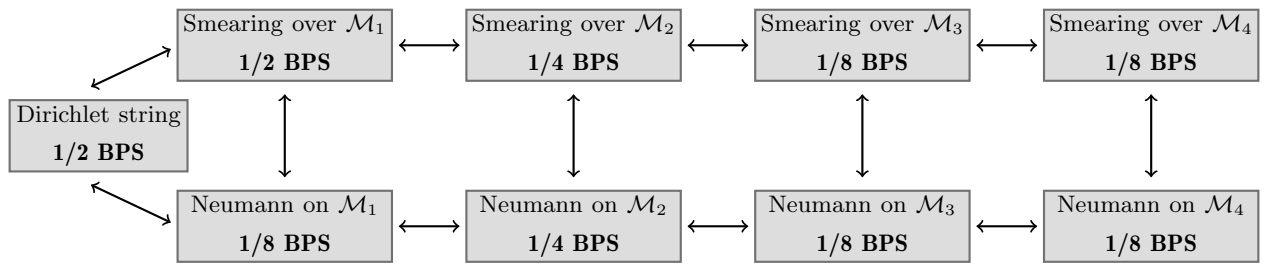

\section{Conclusions}

As we have argued in this paper, there must exist a rich family of line operators in the 2-dimensional CFTs dual to string theories on  $AdS_3\times S^3\times S^3\times S^1$.  The knowledge of their symmetries and supersymmetries can be useful to account for correlators in the superconformal lines and, from them, to infer physical information about the CFT$_2$ themselves.

The richness of BPS lines is associated with the diversity of supersymmetric open string boundary conditions that can be implemented in the holographic formulation: as shown, either Dirichlet or a wide variety of delocalized boundary conditions can be imposed in the compact space $S^3\times S^3\times S^1$ while still preserving supersymmetry. We have found that Dirichlet conditions account for 1/2 BPS strings, whereas delocalized strings range from 1/2 BPS to 1/8 BPS. In the case of delocalized strings, we have found that both smeared (i.e. fixing $\dot{\alpha}$ for the scalar fields) or Neumann (i.e. fixing $\beta$) boundary conditions can be imposed preserving some amount of supersymmetry. In contrast to what was previously found in other backgrounds such as $AdS_5 \times S^5$ or $AdS_4 \times CP^3$, we have found cases in which smeared and Neumann boundary conditions along a same direction do not preserve the same amount of supersymmetry, and cases in which strings with smeared boundary conditions preserve the same number of supersymmetries as in the Dirichlet string. Finally, we have found that all the supersymmetric strings described are connected between each other by a network of interpolating BPS boundary conditions.  

The fact that worldsheets ending along lines at the boundary can be supersymmetric for boundary conditions other than Dirichlet is due to the existence of massless fermionic excitations \cite{Correa:2019rdk}. This is somehow reminiscent to the richness of the protected operators in spectrum of integrable AdS$_3$/CFT$_2$ backgrounds
\cite{Babichenko:2009dk,OhlssonSax:2011ms,Cagnazzo:2012se}, whose origin is also attributed to the existence of massless fermionic excitations in the integrable description \cite{OhlssonSax:2012smh}. It would be interesting to further investigate the relation between these two properties.

 A further characterization of the line operators described in the above sections could be achieved by computing correlations between the components of the displacement multiplet, which are in correspondence with the fluctuations around classical worldsheet configurations. This could be done by extending the analysis of \cite{Beccaria:2017rbe,Beccaria:2019dws,Bianchi:2020hsz} to the open strings presented in this paper for the $AdS_3\times S^3\times S^3\times S^1$ background.
The displacement multiplet constitutes a particular representation of the supergroup of symmetries of the line defect. Correspondingly, excitations on the worldsheet should fit into representations of the symmetry group of the classical string. The symmetry group of the type IIB theory in $AdS_3 \times S^3 \times S^3 \times S^1$ is the direct product $D(2,1;\sin^2 \Omega) \otimes D(2,1;\sin^2 \Omega)$, where $D(2,1;\sin^2 \Omega)$ is an exceptional supergroup whose bosonic subalgebra is $\mathfrak{sl}(2,\mathbb{R}) \oplus \mathfrak{su}(2) \oplus \mathfrak{su}(2)$ \cite{Babichenko:2009dk,Gauntlett:1998kc}.
This suggests that the classical string described in this paper is invariant under a single $D(2,1;\sin^2 \Omega)$,
and the fluctuations in the case of Dirichlet boundary conditions should form a representation of this supergroup. Similarly, fluctuations in the case of smeared or Neumann boundary conditions should transform in representations of smaller subgroups of $D(2,1;\sin^2 \Omega)$. The existence of a rich moduli of BPS line operators in the dual CFTs allows for many future non-trivial comparisons 
between different
computations of correlators of 
displacement multiplet components
by employing superconformal bootstrap techniques in the CFT side or by using Witten diagram on the worldsheet, as in \cite{Liendo:2018ukf,Bianchi:2020hsz}.

\section*{Acknowledgements}
This work was supported by PIP 0688,  PIP UE 084 {\it B\'usqueda de Nueva F\'isica},  UNLP X850 and UNLP X910.
M. Lagares is supported by a fellowship from CONICET.

\appendix

\section{ $AdS_3\times S^3\times S^3\times S^1$ Killing Spinors}
\label{killingspinors}
For the coordinates \eqref{poincare}-\eqref{spherecoor} and the choice of vielbeins
\begin{empheq}{alignat=9}
\label{vierbein1}
    e^0 & = \frac{L}{z}dt\,, &\qquad  e^1 & = \frac{L}{z}dx \,, &\qquad & e^2 = \frac{L}{z}dz\,,
    \\
e^3 & = \frac{L}{\sin \Omega}d\beta_+\,, &\qquad  e^4 & = \frac{L}{\sin \Omega}\cos\beta_+d\gamma_+ \,,&\qquad & e^5 = \frac{L}{\sin \Omega}\cos\beta_+\cos\gamma_+d\varphi_+\,,
    \\
 e^6 & = \frac{L}{\cos \Omega}d\beta_-\,, &\qquad  e^7 & = \frac{L}{\cos \Omega}\cos\beta_-d\gamma_- \,,&\qquad & e^8 = \frac{L}{\cos \Omega}\cos\beta_-\cos\gamma_-d\varphi_-\,,
   \\
 e^9 & = l d\theta\,, & & & & & &
 \label{vierbein4}
\end{empheq}
the non-vanishing components of the spin connections turn out to be
\begin{empheq}{alignat=9}
    \omega^{20} & = \frac{1}{z}dt\,, &\qquad  \omega^{21} & = \frac{1}{z}dx \,, &\qquad &     
    \\
\omega^{34} & = \sin\beta_+d\gamma_+\,, &\qquad  \omega^{35} & = \sin\beta_+\cos\gamma_+d\varphi_+ \,,&\qquad & \omega^{45} = \sin\gamma_+d\varphi_+\,,
    \\
    \omega^{67} & = \sin\beta_-d\gamma_-\,, &\qquad  \omega^{68} & = \sin\beta_-\cos\gamma_-d\varphi_- \,,&\qquad & \omega^{78} = \sin\gamma_-d\varphi_-\,.
\end{empheq}

The conditions defining the Killing spinors can be conveniently presented in terms of a complex 3-form field strength:
\begin{equation}
G_{(3)} = - e^{-\Phi/2} H_{(3)} - i e^{\Phi/2} F_{(3)}
 = 2 i L^2 e^{-\Phi/2+i\lambda}
\left({\rm vol}(AdS_3)+\frac{1}{\sin^2 \Omega} {\rm vol}(S^3_+)+ \frac{1}{\cos^2 \Omega} {\rm vol}(S^3_-) \right)
\,.   
\end{equation}

More precisely, for a real representation of Dirac matrices $\Gamma_\mu = e^m_\mu \gamma_m$, a Killing spinor $\epsilon$ satisfies
\begin{align}
\Gamma^{\mu\nu\rho} G_{\mu\nu\rho}   \epsilon = & 0\,,
\label{killingcond1}
\\
D_\mu\epsilon +\frac{e^{\Phi/2}}{96}\left(9\Gamma^{\nu\rho}G_{\mu\nu\rho}  - \Gamma_\mu^{\ \, \nu\rho\sigma}G_{\nu\rho\sigma}\right)\epsilon^{*} = & 0\,,
\label{killingcond2}
\end{align}
where $D_\mu$ stands for the covariant derivative
\begin{equation}
D_\mu = \partial_\mu +\frac{1}{4} \omega^{mn}_\mu
\gamma_{mn}\,.
\end{equation}
Additionally, the Killing spinors are Weyl spinors obeying
\begin{equation}
\gamma_{11}\epsilon = -\epsilon\,,
\end{equation}
where $\gamma_{11} = \gamma^0\gamma^1\cdots\gamma^9$. Defining
\begin{equation}
\gamma_* =   \gamma^0\gamma^1\gamma^2\,,
\qquad
\gamma_*^+ =  i \gamma^3\gamma^4\gamma^5\,,
\qquad
\gamma_*^- =   i \gamma^6\gamma^7\gamma^8\,,
\end{equation}
and
\begin{equation}
    \label{sigmaprojector}
    P^\Sigma_{\pm} :=  \frac12(1 \pm \Sigma) \qquad \text{for}\qquad
     \Sigma := i\left(\sin \Omega \, \gamma_*\gamma_*^++\cos \Omega \, \gamma_*\gamma_*^-\right) \,
\end{equation}
the condition \eqref{killingcond1} becomes
\begin{equation}
   P^\Sigma_{-}\epsilon =0\,.
    \label{killingcond1bis}
\end{equation}

Since the condition \eqref{killingcond2} relates $\epsilon$ with $\epsilon^*$  it is convenient to separate the Killing spinor into
\begin{equation}
\epsilon = \eta + i \xi\,.
\label{etaxi}
\end{equation}
For the different values of $\mu$, and using the condition \eqref{killingcond1bis}, the equation \eqref{killingcond2}  becomes
\begin{empheq}{alignat=9}
\label{killeq1}
   \partial_t  \eta  -  \frac{1}{z} M_x \eta 
   -\frac{1}{z} \sin\lambda M_t \eta +\frac{1}{z} \cos\lambda M_t \xi
   & =0\,, &\qquad  \partial_t  \xi  -  \frac{1}{z} M_x \xi 
   +\frac{1}{z} \sin\lambda M_t \xi +\frac{1}{z} \cos\lambda M_t \eta
   & =0\,,&     
    \\
    \label{killeq2}
    \partial_x  \eta  -  \frac{1}{z} M_t \eta 
   -\frac{1}{z} \sin\lambda M_x \eta -\frac{1}{z} \cos\lambda M_x \xi
   & =0\,, &\qquad  \partial_x  \xi  -  \frac{1}{z} M_t \xi 
   +\frac{1}{z} \sin\lambda M_x \xi -\frac{1}{z} \cos\lambda M_x \eta
   & =0\,,&     
   \\
   \partial_z  \eta   
   -\frac{1}{z} \sin\lambda M_z \eta +\frac{1}{z} \cos\lambda M_z \xi
   & =0\,, &\qquad  \partial_z  \xi  
   +\frac{1}{z} \sin\lambda M_z \xi +\frac{1}{z} \cos\lambda M_z \eta
   & =0\,,&  
   \label{killeq4}
   \\
   \label{killeq5}
   \partial_{\theta}  \eta   & =0\,, 
   &\qquad  
   \partial_{\theta}  \xi  & =0\,,&    
 \\
 \label{killeq6}
   \partial_{\beta_\pm}  \eta   
   -\sin\lambda M_{\beta_\pm} \eta +\cos\lambda M_{\beta_\pm} \xi
   & =0\,, &\qquad  \partial_{\beta_\pm}  \xi 
   + \sin\lambda M_{\beta_\pm} \xi + \cos\lambda M_{\beta_\pm} \eta
   & =0\,,&  
   \end{empheq}
   \vspace{-1.25cm}
   
   \begin{empheq}{alignat=9}
   \label{killeq7}
  & & \partial_{\gamma\pm}  \eta   +\sin\beta_{\pm} M_{\varphi_{\pm}}\eta
   -\sin\lambda \cos\beta_{\pm} M_{\gamma_\pm} \eta +\cos\lambda \cos\beta_{\pm} M_{\gamma_\pm} \xi
   & =0\,, &
   \\
   \label{killeq8}
  & & \partial_{\gamma\pm}   \xi +\sin\beta_{\pm} M_{\varphi_{\pm}}\xi 
 +\sin\lambda \cos\beta_{\pm} M_{\gamma_\pm} \xi + \cos\lambda \cos\beta_{\pm} M_{\gamma_\pm} \eta
   & =0\,,& 
   \\
   \label{killeq9}
    & & \partial_{\varphi\pm}  \eta   -\sin\beta_{\pm}\cos\gamma_{\pm} M_{\gamma_{\pm}}\eta
    +\sin\gamma_{\pm} M_{\beta_{\pm}}\eta
   -\sin\lambda \cos\beta_{\pm}\cos\gamma_{\pm} M_{\varphi_\pm} \eta +\cos\lambda \cos\beta_{\pm}\cos\gamma_{\pm} M_{\varphi_\pm} \xi
   & =0\,, &
   \\
   \label{killeq10}
       & & \partial_{\varphi\pm}  \xi   -\sin\beta_{\pm}\cos\gamma_{\pm} M_{\gamma_{\pm}}\xi
    +\sin\gamma_{\pm} M_{\beta_{\pm}}\xi
   +\sin\lambda \cos\beta_{\pm}\cos\gamma_{\pm} M_{\varphi_\pm} \xi +\cos\lambda \cos\beta_{\pm}\cos\gamma_{\pm} M_{\varphi_\pm} \eta
   & =0\,, &
   \end{empheq}
where the matrices $M$ are the ones defined in \eqref{Mtxz}-\eqref{Mminus}.

By solving the Killing equations \eqref{killeq1}-\eqref{killeq10} one by one, we will construct a factorized expression for the Killing spinors. We can start with equations \eqref{killeq5}, which simply state that the Killing spinors are independent of $\theta$. Next, we consider the equations \eqref{killeq6}. They can be disentangled by taking an additional derivative with respect to $\beta_\pm$, which gives
\begin{equation}
\partial_{\beta_\pm}^2 \eta = M_{\beta_\pm}^2 \eta\,,\qquad   
\partial_{\beta_\pm}^2 \xi = M_{\beta_\pm}^2 \xi\,.
\label{kill5derivated}
\end{equation}
Let us first solve for the $\beta_+$ dependence. In order to fulfill \eqref{kill5derivated} we need
\begin{align}
\eta & = e^{\beta_+ M_{\beta_+}} a + e^{-\beta_+ M_{\beta_+}} b \,,
\\
\xi & = e^{\beta_+ M_{\beta_+}} c + e^{-\beta_+ M_{\beta_+}} d 
\,.
\end{align}
Replacing in \eqref{killeq6} we find a relation between $a$ and $c$ and between $b$ and $d$. Defining $U_2 = e^{\beta_+ M_{\beta_+}}$ and $V_2 = e^{-\beta_+ M_{\beta_+}}$:
\begin{align}
\eta & = U_2 \epsilon_0^{(2)} + \frac{\cos\lambda}{\sin\lambda +1} V_2 \epsilon_1^{(2)} \,,
\\
\xi & =-\frac{\cos\lambda}{\sin\lambda +1} U_2 \epsilon_0^{(2)} + V_2 \epsilon_1^{(2)} 
\,,
\end{align}
where $\epsilon_0^{(2)}$ and $\epsilon_1^{(2)}$ are not constant but independent of $\theta$ and $\beta_+$. In the same way we can deal with the $\beta_-$ dependence. Now we define 
$U_3 = e^{\beta_+ M_{\beta_+}}e^{\beta_- M_{\beta_-}}$ and $V_3 = e^{-\beta_+ M_{\beta_+}}e^{-\beta_- M_{\beta_-}}$ and then
\begin{align}
\eta & = U_3 \epsilon_0^{(3)} + \frac{\cos\lambda}{\sin\lambda +1} V_3 \epsilon_1^{(3)} \,,
\\
\xi & =-\frac{\cos\lambda}{\sin\lambda +1} U_3 \epsilon_0^{(3)} + V_3 \epsilon_1^{(3)} 
\,.
\end{align}
It is straightforward to extend this to account completely for the dependence on the compact space coordinates
\begin{empheq}{alignat=9}
\eta & = U_7 \epsilon_0^{(7)}(t,x,z) + \frac{\cos\lambda}{\sin\lambda +1} V_7 \epsilon_1^{(7)}(t,x,z) \,,
\\
\xi & =-\frac{\cos\lambda}{\sin\lambda +1} U_7 \epsilon_0^{(7)}(t,x,z) + V_7 \epsilon_1^{(7)}(t,x,z) 
\,,
\end{empheq}
where
\begin{align}
\label{UyV}
 U_7 & = e^{\beta_+ M_{\beta_+}+\beta_- M_{\beta_-}}
 e^{\gamma_+ M_{\gamma_+} +\gamma_- M_{\gamma_-}}
 e^{\varphi_+ M_{\varphi_+}+\varphi_- M_{\varphi_-}}\,, \\
 V_7 & = e^{-\beta_+ M_{\beta_+}-\beta_- M_{\beta_-}}
 e^{-\gamma_+ M_{\gamma_+}-\gamma_- M_{\gamma_-}}
 e^{-\varphi_+ M_{\varphi_+}-\varphi_- M_{\varphi_-}}\,.
\end{align}

Finally, one can solve for the remaining eqs. \eqref{killeq1}-\eqref{killeq4} to obtain
\begin{empheq}{alignat=9}
\label{eps07}
\epsilon_0^{(7)}(t,x,z)  & = e^{\log z M_z}e^{(x+t)(M_t+M_x)}\epsilon_0 
= \frac12 \left[\frac{1}{\sqrt{z}}(1-2M_z)\left(1+2(x+t)M_t\right)+\sqrt{z}(1+2M_z)\right]\epsilon_0
\,, 
\\
\label{eps17}
\epsilon_1^{(7)}(t,x,z)  & = e^{-\log z M_z}e^{(x-t)(M_t-M_x)}\epsilon_1 
= \frac12 \left[\frac{1}{\sqrt{z}}(1+2M_z)\left(1+2(x-t)M_t\right)+\sqrt{z}(1-2M_z)\right]\epsilon_1
\,.
\end{empheq}

\begin{section}{Quadratic fluctuations}
\label{mass spectrum apendix}
Before presenting the mass spectrum for the quadratic fluctuations around \eqref{susyconfiguration}, let us see that it also emerges as a solution to the string equations of motion. The action for the bosonic coordinates of the string coupled to the NS-NS $B$-field is
\begin{equation}
\label{bosonic action}
S_B=-\int d^2 \sigma \; \sqrt{-h} + \frac{1}{2} \int d^2 \sigma \; \epsilon^{\alpha \beta} B_{\alpha\beta}\,.
\end{equation}
Here $h_{\alpha\beta}$ and $B_{\alpha\beta}$ are the induced metric and the pullback of the NS-NS $B$-field on the worldsheet, respectively. For the $B$-field we will use the gauge
\begin{equation}
    \label{b field gauge}
    B= 2 L^2 \sin \lambda \left( \frac{t}{z^3} \; dx\wedge dz+ \frac{\varphi_+}{\sin^2 \Omega} \; \cos^2 \beta_+ \; \cos \gamma_+ \; d\beta_+ \wedge d\gamma_+ + \frac{\varphi_-}{\cos^2 \Omega} \; \cos^2 \beta_- \; \cos \gamma_- \; d\beta_- \wedge d\gamma_- \right)\,.
\end{equation}
Making the ansatz 
\begin{equation}
t = \omega\tau\,,\qquad
x = x(\sigma)\,,\qquad 
z = \sigma\,,\qquad 
\beta_\pm,\gamma_\pm,\varphi_\pm,\theta = \text{const.}
\label{ansatzapp}
\end{equation}
the equations of motions reduce to
\begin{equation}
2\left(1+(x')^2\right)\left(
x' +\sin\lambda \sqrt{1+(x')^2}\right)
-\sigma x'' = 0\,,
\end{equation}
which is straightforwardly solved by $x' = -\tan\lambda$.

Let us now turn to the mass spectrum of the bosonic fluctuations. The standard approach would be to perform a series expansion of the action \eqref{bosonic action} in powers of $\delta X^{\mu}:={X}^{\mu}-X^{\mu}_{\sf cl}$, with $X^{\mu}_{\sf cl}$ the classical solution.
However, and because the coefficients of such expansion will not be manifestly covariant \cite{Drukker:2000ep,Forini:2015mca}, it is more convenient to make the expansion in terms of Riemann normal coordinates. Let $X^{\mu}(q)$ be a geodesic such that $X^{\mu}(0)=X^{\mu}_{\sf cl}$. Using the geodesic equation we get
\begin{equation}
    \label{geodesic eq solution}
    X^{\mu}(q)=X^{\mu}_{\sf cl}+q \zeta^{\mu}-\frac{q^2}{2} \Gamma^{\mu}_{\nu \sigma} (X^{\mu}_{\sf cl}) \zeta^{\nu} \zeta^{\sigma} + \mathcal{O}(q^3)\,,
\end{equation}
where we have defined
\begin{equation}
    \label{rho coord riemann}
    \zeta^{\mu}:=\left.\frac{dX^{\mu}}{dq}\right|_{q=0}\,.
\end{equation}
If we now look for a geodesic with $X^{\mu}(1)=\tilde{X}^{\mu}$, we then have
\begin{equation}
    \label{rho expansion}
    \tilde{X}^{\mu}=X^{\mu}_{cl}+ \zeta^{\mu}-\frac{1}{2} \Gamma^{\mu}_{\nu \sigma} (X^{\mu}_{\sf cl}) \zeta^{\nu} \zeta^{\sigma} + \mathcal{O}(\zeta^3)\,,
\end{equation}
which serves as an expansion of $\tilde{X}^{\mu}$ around $X^{\mu}_{\sf cl}$ in powers of a vector $\zeta^{\mu}$. Moreover, in order to get an expansion in terms of scalar fields, it is useful to define 
\begin{equation}
    \label{scalar phi}
    \phi^m= e^{m}_{\mu} \zeta^{\mu}\,,
\end{equation}
where $e^a_{\mu}$ are the inverse of the background-space vielbeins given in \eqref{vierbein1}-\eqref{vierbein4}. 

Plugging \eqref{rho expansion} into \eqref{bosonic action} we get, up to some boundary terms which can be discarded by the addition of counterterms, 
\begin{equation}
\label{bosonic action expansion}
S_B=S_{\sf cl}-\frac{1}{2} \int d^2 \sigma \; \sqrt{-h_{\sf cl}} \left[ \partial_{\alpha} \phi^{\sf tr} \partial^{\alpha} \phi^{\sf tr} + \frac{2}{R^2} (\phi^{\sf tr})^2 + \sum_{j=3}^9 \left(  \partial_{\alpha} \phi^{j} \partial^{\alpha} \phi^{j} \right) \right]\,,
\end{equation}
where $S_{\sf cl}$ and $(h_{\alpha\beta})_{\sf cl}$ stand respectively for the action and the induced metric evaluated in the classical solution (in what follows we will refer to $(h_{\alpha\beta})_{\sf cl}$ simply as $h_{\alpha\beta}$), and $R=L/\cos \lambda$ is the radius of the induced $AdS_2$. $\phi^{\sf tr}$ stands for the 
transverse fluctuation in $AdS_3$, defined as
\begin{equation}
    \label{transverse fluctuation}
    \phi^{\sf tr} (\tau,\sigma):= \cos \lambda \; \phi^1(\tau,\sigma) +\sin \lambda\; \phi^2(\tau,\sigma)\,.
\end{equation}
Then, we see from \eqref{bosonic action expansion} that the spectrum of bosonic fluctuations consists in 1 massive (with mass $m_{\sf tr}^2=\frac{2}{R^2}$) and 7 massless scalar fields.

Regarding the fermionic fluctuations $\Theta^1$ and $\Theta^2$, and in order to simplify the calculations, we will fix the $\kappa$-symmetry gauge condition to be $\Theta^1=\Theta^2:=\Theta$, as customary for open strings in type IIB backgrounds \cite{Drukker:2000ep}. Then, up to quadratic order in the fluctuations the action reads \cite{Cvetic:1999zs}
\begin{equation}
    \label{fermionic action}
    S_F= S_{F1}+S_{F2}+S_{F3}\,,
\end{equation}
with 
\begin{align}
    \label{SF1}
    S_{F1} &:=-{2 i} 
\int d^2 \sigma \sqrt{-h} \; h^{\alpha\beta} \bar{\Theta} \Gamma_{\alpha}  {\cal D}_{\beta}  \Theta\,,    \\
      \label{SF2}
    S_{F2} &:=-\frac{i e^{\Phi}}{24}  \int d^2 \sigma \sqrt{-h} \; h^{\alpha\beta}  \; \bar{\Theta} \Gamma_{\alpha}  \Gamma^{\delta \rho \eta} F_{\delta \rho \eta} \Gamma_{\beta} \Theta = 
    \frac{i}{R} \int d^2 \sigma \sqrt{-\gamma} \; \bar{\Theta} \gamma_* P_{\Sigma} \Theta
    \,, \\
     \label{SF3}
    S_{F3} &:=-\frac{i}{2} \int d^2 \sigma  \; \epsilon^{\alpha\beta} \partial_{\alpha} X^{\mu} \partial_{\beta} X^{\nu} \; \bar{\Theta} \Gamma_{\mu}\;^{\rho \eta} H_{\nu\rho\eta}  \Theta = 0\,,
\end{align}
where 
\begin{equation}
    \label{string covariant derivative}
\Gamma_{\alpha}:= \partial_{\alpha} X^{\mu} \Gamma_{\mu}\,, \qquad
{\cal D}_{\alpha}:=\partial_{\alpha} + \frac{1}{4} \tilde\omega_{\alpha}^{ab} {\tilde\gamma}_{ab}\,,
\end{equation}
with ${\tilde\gamma}^{{a}}:=\tilde{e}^{{a}}_{\alpha} \Gamma^{\alpha}$, for $\tilde{e}^{{a}}_{\alpha}$ and $\tilde\omega_{\alpha}^{ab}$ the vielbeins and spin connections of $AdS_2$. Further splitting $\Theta$ as
\begin{equation}
    \Theta = \Theta_0 +\Theta_1\,, \qquad\text{for}\qquad
P^\Sigma_-\Theta = \Theta_0\,,\quad P^\Sigma_+\Theta =\Theta_1\,,
\label{Sigmaprojections}
\end{equation}
the total action for the fermionic fluctuations becomes
\begin{equation}
    \label{fermionic action-v2}
    S_F= -2i \int d^2 \sigma \sqrt{-h} \;\bar{\Theta}_0  \Gamma^{\alpha}  {\cal D}_{\alpha}  \Theta_0+ \bar{\Theta}_1 \left( \Gamma^{\alpha}  {\cal D}_{\alpha} -\frac{1}{R} \gamma_* \right) \Theta_1\,.
\end{equation}

\begin{subsection}{Supersymmetry of the fluctuations}
\label{susyflucapp}

The Green-Schwarz action is invariant under supersymmetry transformations
\begin{equation}
\delta_{\sf susy} X^\mu = \bar\eta \Gamma^\mu \Theta^{1}
+\bar\xi \Gamma^\mu \Theta^{2}
\,,\qquad
\delta_{\sf susy} \Theta^1 = \eta
\,,\qquad
\delta_{\sf susy} \Theta^2 = \xi\,,
\end{equation}
for $\eta$ and $\xi$ the real and imaginary parts of the Killing spinors \eqref{etaxi}. It is also invariant under the $\kappa$-symmetry transformations
\begin{equation}
\delta_{\kappa} X^\mu = \bar\Theta^1\Gamma^\mu P^\kappa_+\kappa^1
+\bar\Theta^2 \Gamma^\mu P^\kappa_-\kappa^2
\,,\qquad
\delta_{\kappa} \Theta^1 = P^\kappa_+\kappa^1
\,,\qquad
\delta_{\kappa} \Theta^2 = P^\kappa_-\kappa^2\,,
\end{equation}
where the projectors
\begin{equation}
P^\kappa_\pm = \frac12\left(1\pm \tilde\Gamma\right) \,,
\qquad 
\tilde\Gamma = -\frac{\epsilon^{\alpha\beta}\Pi^\mu_\alpha\Pi^\nu_\beta\Gamma_{\mu\nu}}{2\sqrt{-H}}\,,
\end{equation}
are defined in terms of
\begin{equation}
\Pi^\mu_\alpha = \partial_\alpha X^\mu -  \bar\Theta^1 \Gamma^\mu \partial_\alpha\Theta^{1}-  \bar\Theta^2 \Gamma^\mu \partial_\alpha\Theta^{2}\,,
\qquad
H_{\alpha\beta} = g_{\mu\nu} \Pi^\mu_\alpha \Pi^\nu_\beta\,.
\end{equation}

Whenever a combined transformation
$\delta:=\delta{\sf susy}+\delta_\kappa$ leaves a configuration invariant, we say it is supersymmetric. A classical configuration with vanishing fermions is then supersymmetric for $\kappa_1 = -\eta$ and $\kappa_2 = -\xi$ if
\begin{equation}
 P^\kappa_- \eta = 0\,, \qquad P^\kappa_+ \xi = 0\,,
 \qquad \Leftrightarrow \qquad 
\tilde\Gamma K \epsilon = \epsilon\,,
\label{kappapro}
\end{equation}
where $K$ indicates complex conjugation.

As we have shown, the string configuration \eqref{susyconfiguration} is supersymmetric. In what follows we shall derive the supersymmetry transformations for the fluctuations around it. The preservation of the $\kappa$-symmetry gauge condition requires $\delta \Theta^1 = \delta \Theta^2$ and from this (labelling with $(0), (1), \ldots $ the successive orders of the expansion in powers of the fluctuations), we get
\begin{equation}
\eta_{(1)} + P_{+(0)}^{\kappa} \kappa^1_{(1)} -  P_{+(1)}^{\kappa} \eta_{(0)}  =
\xi_{(1)} + P_{ -(0)}^{\kappa} \kappa^2_{(1)} -  P_{ -(1)}^{\kappa}  \xi_{(0)}\,.
\end{equation}

For the variation of the bosonic transverse fluctuations we obtain\footnote{The total transformation must preserve the static gauge choice, which requires $\delta X^0=\delta X^{\sf lg}=0$. In order to achieve this, a diffeomorphism should be added to the total transformation $\delta$. 
As the transformation should be first order in the fluctuations, $\delta_{\sf diff} X^{\mu}=-\partial_{\alpha} X^{\mu}_{\sf cl} \delta \sigma^{\alpha}$. Then, this additional transformation does not affect the transverse fluctuations and $\delta\sigma^\alpha$ can be chosen so that $\delta X^0=\delta X^{\sf lg}=0$. Similarly, $\delta_{\sf diff} \Theta=-\partial_{\alpha} \Theta_{\sf cl} \, \delta \sigma^{\alpha}=0$ and so the diffeomorphism does not modifies the transformation of the fermionic fluctuations either. }
\begin{align}
\label{susy bosonic fluctuations}
\delta\phi^{\sf tr} &=    -2 \bar\Theta \gamma^{\sf tr}\left(\eta_{(0)}+\xi_{(0)}\right)\,, \\
\delta\phi^a &=    -2 \bar\Theta \gamma^a\left(\eta_{(0)}+\xi_{(0)}\right)\,, \qquad a=3, \dots, 9
\end{align}
while for the fermionic fluctuations we get
\begin{equation}
\delta\Theta = \frac{1}{2}\left(\eta_{(1)}+ \xi_{(1)} \right) -
\frac{1}{2}\tilde\Gamma_{(0)}\left(\eta_{(1)} - \xi_{(1)} \right)-
\frac{1}{2}\tilde\Gamma_{(1)}\tilde\Gamma_{(0)}\left(\eta_{(0)} + \xi_{(0)} \right)\,,
\label{deltaTheta}
\end{equation}
where
\begin{align}
\tilde\Gamma_{(0)} & = -2\sin\lambda M_z -2 \cos\lambda M_x := -2 M_{\sf tr}\,, 
\\
\tilde\Gamma_{(1)} & = -2\left[ 
\frac{\sigma}R \partial_\tau \phi^{\sf tr} M_t -\frac1{R}\left(\phi^{\sf tr}
+\sigma\partial_\sigma \phi^{\sf tr}\right)M_{\sf lg}
+\frac{\sigma}{2R} \sum_{a=3}^9\gamma^a
\left(\partial_\tau\phi^a \gamma^{\sf lg}+\partial_\sigma\phi^a \gamma^{0}\right)
\right]\,,
\label{Gamma1a}
\end{align}
with
\begin{equation}
M_{\sf lg} = \cos\lambda M_z - \sin\lambda M_x \,.     
\end{equation}
For getting \eqref{deltaTheta} we have used that
\begin{equation}
\eta_{(0)} - \xi_{(0)} = \tilde\Gamma_{(0)}\left(\eta_{(0)} + \xi_{(0)} \right) \,.
\end{equation}

Using this relation, it is also possible to express all the dependence of $\delta\Theta$ on the Killing spinors through $\eta_{(0)} + \xi_{(0)}$, since
\small{
\begin{align}
\eta_{(1)} + \xi_{(1)} =  &  -\frac{1}{L}\left[\left(\phi^{\sf tr}M_{\sf tr}+
\sin \Omega\sum_{a=3}^5\phi^a M_a+\cos \Omega\sum_{a=6}^8\phi^a M_a
\right)(\cos\lambda-\sin\lambda\tilde\Gamma_{(0)})-\cos\lambda \phi^{\sf tr} M_t
\right](\eta_{(0)} + \xi_{(0)})\,, \nonumber
\\
\eta_{(1)} - \xi_{(1)} =  &
 \frac{1}{L}\left[\left(\phi^{\sf tr}M_{\sf tr}+
\sin \Omega\sum_{a=3}^5\phi^a M_a+\cos \Omega\sum_{a=6}^8\phi^a M_a
\right)(\sin\lambda+\cos\lambda\tilde\Gamma_{(0)})
+\cos\lambda \phi^{\sf tr} M_t \tilde{\Gamma}_{(0)}
\right](\eta_{(0)} + \xi_{(0)})\,, \nonumber
\end{align}
}
\normalsize{}
Thus, splitting into the variation of massless and massive fermions, taking into account that $\Theta_0 =P^\Sigma_-\Theta$,  $\Theta_1 =P^\Sigma_+\Theta$ and that $P^\Sigma_-(\eta_{(0)} + \xi_{(0)}) = 0$, we have
\small{
\begin{align}
\delta\Theta_0 = &
\left(-\cos \Omega\sum_{a=3}^5\Gamma^\alpha \partial_\alpha\phi^a M_a+\sin \Omega\sum_{a=6}^8 \Gamma^\alpha \partial_\alpha\phi^a M_a
+\tfrac{1}{2}\Gamma^\alpha \partial_\alpha\phi^9\right)\gamma^9(\eta_{(0)} + \xi_{(0)})\,,
\\
\delta\Theta_1 = & \left(\tfrac{1}{2}\Gamma^\alpha\partial_\alpha\phi^{\sf tr}\gamma^{\sf tr}
-\tfrac{1}{R}\phi^{\sf tr}M_{\sf tr}
+
\sin \Omega\sum_{a=3}^5\left(\Gamma^\alpha \partial_\alpha\phi^a\gamma_*-\tfrac{1}{R}\phi^a \right)M_a  
+\cos \Omega\sum_{a=6}^8\left(\Gamma^\alpha \partial_\alpha\phi^a\gamma_*-\tfrac{1}{R}\phi^a \right)M_a
\right)(\eta_{(0)} + \xi_{(0)}). \nonumber
\end{align}
}

\normalsize{}

\end{subsection}

\begin{subsection}{Fluctuations as $AdS_2$ fields}
\label{ads2 fields appendix}

In the previous subsection, fermionic fluctuations and Killing fields are 32-component spinors. We can alternatively describe the quadratic fluctuations and their supersymmetry transformations in terms of 2-components $AdS_2$ spinors. In order to do that, it is necessary to identify longitudinal and transverse directions (to the worldsheet) and decompose the Dirac matrices into products are $SO(1,1)$ and $SO(8)$ matrices:
\begin{align}
\gamma^0    & = \uptau^0\otimes \mathds{1}_{16}\,,
\\
\gamma^{\sf lg} & = -\sin\lambda \gamma^1+\cos\lambda \gamma^2 =
\uptau^1\otimes \mathds{1}_{16}\,,
\\
\gamma^{\sf tr}  & =  \cos\lambda \gamma^1+\sin\lambda \gamma^2 =
\uptau_3\otimes \uprho^8\,,
\\
\gamma^{m}  & = \uptau_3\otimes \uprho^{m-2}\,,\qquad\text{for } m\geq 3\,,
\end{align}
where $\uptau^a$ and $\uprho^A$ are $SO(1,1)$ and $SO(8)$ Dirac matrices. Then, the 32-component spinors can be written as $\Theta = \psi(\tau,\sigma)\otimes\varphi$, where $\psi(\tau,\sigma)$ are 2-component worldsheet spinors and $\varphi$ are constant 16-component spinors. All these spinors can be accommodated into representations of the $SU(2)$ rotations that leave invariant the $\Sigma$ projections \eqref{Sigmaprojections}. Thus,
\begin{equation}
\Theta_0 = \psi_{+}^{ a\dot{a}}\otimes  \varphi_{{ a\dot{a}}}^{++}+\psi_{-}^{a\dot{a}}\otimes  \varphi_{{ a\dot{a}}}^{--}\,,\qquad
\Theta_1 = \chi_{+}^{ a\dot{a}}\otimes  \varphi_{{ a\dot{a}}}^{+-} +\chi_{-}^{ a\dot{a}}\otimes  \varphi_{{ a\dot{a}}}^{-+}\,,\qquad
\end{equation}
where ${ a}$ and $\dot{ a}$ are fundamental $SU(2)$ indices and
\begin{align}
\uptau_3\psi_\pm = \pm \psi_\pm\,,\qquad
    \uprho_9\varphi^{\pm s} = \pm \varphi^{\pm s}\,,\qquad 
    \upsigma\varphi^{s\pm} = \pm \varphi^{s\pm}\,,
\end{align}
for
\begin{equation}
\uptau_3 = \uptau^0\uptau^1\,,\qquad 
\uprho_9 = \uprho^1\cdots\uprho^8\,,\qquad
\upsigma = \sin \Omega \ \uprho^8\uprho^1\uprho^2\uprho^3+\cos \Omega \ \uprho^8\uprho^4\uprho^5\uprho^6\,.
\end{equation}
It is easy and convenient to use a basis
\footnote{In the need of an explicit basis we shall use
\begin{align}
    \uptau^0 = i\sigma^2\,,\qquad \uptau^1 = \sigma^1 \,,
    \nonumber
\end{align}
and
\begin{empheq}{alignat=9}
    \uprho^1 &= \sigma^2\otimes\sigma^2\otimes\sigma^1\otimes \mathds{1}_{2}\,, &\qquad 
    \uprho^2 &= \sigma^2\otimes\sigma^2\otimes\sigma^3\otimes \mathds{1}_{2}\,, &\qquad
    \uprho^3 &= \sigma^2\otimes\sigma^1\otimes \mathds{1}_{2} \otimes\sigma^2\,, &\qquad 
    \uprho^4 &= \sigma^2\otimes\sigma^3\otimes \mathds{1}_{2} \otimes\sigma^2\,,
    \nonumber\\
\uprho^5 &= \sigma^2\otimes \mathds{1}_{2}\otimes\sigma^2\otimes\sigma^1\,,  &\qquad 
\uprho^6 &= \sigma^2\otimes \mathds{1}_{2}\otimes\sigma^2\otimes\sigma^3\,,  &\qquad
\uprho^7 &= \sigma^1\otimes\mathds{1}_{2}\otimes\mathds{1}_{2}\otimes \mathds{1}_{2}\,,  &\qquad 
\uprho^8 &= -\sigma^3\otimes\mathds{1}_{2}\otimes\mathds{1}_{2}\otimes \mathds{1}_{2}\,,
    \nonumber
\end{empheq}}
 in which the constant spinors satisfy
\begin{align}
    (\varphi^\dagger)^{{ a\dot{a}}}_{\pm\pm}\cdot\varphi_{{ b\dot{b}}}^{\pm\pm} = \frac14\delta^{ a}_{ b}
    \delta^{ \dot a}_{\dot b}\,,
    \qquad
(\varphi^\dagger)^{{ a\dot{a}}}_{\pm\pm}\cdot\uprho^8\cdot\varphi_{{ b\dot{b}}}^{\mp\mp} = \frac14\delta^{ a}_{ b}
    \delta^{ \dot a}_{ \dot b}\,, \\
    (\varphi^\dagger)^{{ a\dot{a}}}_{\pm\mp}\cdot\varphi_{{ b\dot{b}}}^{\pm\mp} = \frac14\delta^{ a}_{ b}
    \delta^{ \dot a}_{ \dot b}\,,
    \qquad
(\varphi^\dagger)^{{ a\dot{a}}}_{\pm\mp}\cdot\uprho^8\cdot\varphi_{{ b\dot{b}}}^{\mp\pm} = \frac14\delta^{ a}_{ b}
    \delta^{ \dot a}_{ \dot b}\,,
\end{align}
and 0 otherwise.

Substituting in \eqref{fermionic action-v2}, the action for the fermionic fluctuations  becomes
\begin{equation}
    \label{fermionic action-v2d2}
     S_F= -\frac{i}2\! \int\! d^2 \sigma \sqrt{-h} \left(\bar{\psi}_{{ a}\dot{ a}} \ \slash\!\!\!\!{\cal D}^{(2)}  \psi^{{ a}\dot{ a}}  +
    \bar{\chi}_{{ a}\dot{ a}} \left( \slash\!\!\!\!{\cal D}^{(2)}
    -\frac{1}{R}\right)\chi^{{ a}\dot{ a}}\right)\,,
\end{equation}
where
\begin{equation}
   \psi^{{ a}\dot{ a}} = \psi_{+}^{{ a}\dot{ a}}+ \psi_{-}^{{ a}\dot{ a}}\,,
   \qquad
   \chi^{{ a}\dot{a}} = \chi_{+}^{{ a}\dot{ a}}+ \chi_{-}^{{ a}\dot{ a}}\,,
\end{equation}
and $^{(2)}$ indicates that Dirac matrices and covariant derivatives are now those of $AdS_2$. 
This is the action of four massless and four massive (with mass $\frac1R$) worldsheet fermions. 

We would like to also have expressions for the supersymmetry transformations in terms of $AdS_2$ spinors. For this we would need the decomposition of $\eta_{(0)} + \xi_{(0)}$ as well.
 In terms of the matrices $U$ and $V$ \eqref{etadef}-\eqref{xidef} we can write
\begin{equation}
\eta_{(0)} + \xi_{(0)} = \left(1-p(\lambda)\right) U_{(0)}\epsilon_0
+(1+p(\lambda)) V_{(0)}\epsilon_1\,,
\qquad{\rm for}\  p(\lambda) = \frac{\cos\lambda}{\sin\lambda + 1}\,.
\end{equation}
Evaluating \eqref{eps07}-\eqref{eps17} on the classical solutions, using the relation \eqref{susy constraint dirichlet}  and defining
\begin{equation}
    \varepsilon(\tau) = \frac{1}{2(\sin\lambda+1)}\left(1+2M_{\sf lg}\right)\left(1+2M_{\sf tr}\right)(1+2M_z)\left[1+2(x_0-\tau\sec\lambda )M_t\right]\epsilon_1\,,
\end{equation}
we get
\begin{equation}
 \eta_{(0)} + \xi_{(0)} =  \sigma^{-1/2} \varepsilon
 +\sigma^{1/2} 2M_t \dot\varepsilon\,.
\end{equation}
It is useful to note that
\begin{equation}
\label{properties varepsilon}
2M_{\sf lg} \varepsilon = \varepsilon \quad \text{and} \quad \ddot{\varepsilon}=0 \, .
\end{equation}
Then, writing $\eta_{(0)} + \xi_{(0)}$ and $\varepsilon$ as
\begin{equation}
\eta_{(0)} + \xi_{(0)}=\kappa_{+}^{ a\dot{a}}\otimes  \varphi_{{ a\dot{a}}}^{+-} +\kappa_{-}^{ a\dot{a}}\otimes  \varphi_{{ a\dot{a}}}^{-+} \qquad \text{and} \qquad \varepsilon=\varepsilon_{+}^{ a\dot{a}}\otimes  \varphi_{{ a\dot{a}}}^{+-} +\varepsilon_{-}^{ a\dot{a}}\otimes  \varphi_{{ a\dot{a}}}^{-+}
\end{equation}
we obtain
\begin{equation}
 \kappa^{ a\dot{a}} =  \sigma^{-1/2} \varepsilon^{ a\dot{a}} 
 +\sigma^{1/2} \uptau^0 \dot{\varepsilon}^{ a\dot{a}} 
\end{equation}
where
\begin{equation}
   \kappa^{{ a}\dot{ a}} = \kappa_{+}^{{ a}\dot{a}}+ \kappa_{-}^{{ a}\dot{ a}} 
   \qquad \text{and} \qquad
  \varepsilon^{{ a}\dot{a}} = \varepsilon_{+}^{{ a}\dot{ a}}+ \varepsilon_{-}^{{ a}\dot{ a}} \, .
\end{equation}
The properties \eqref{properties varepsilon} can now be expressed as
\begin{equation}
\label{properties varepsilon2}
\uptau^1 \varepsilon^{{ a}\dot{ a}} =- \varepsilon^{{ a}\dot{ a}} \quad \text{and} \quad \ddot{\varepsilon}^{{ a}\dot{ a}}=0 \, .
\end{equation}

Choosing a representation of $\uptau$ and $\uprho$ matrices such that the $\varphi$ spinors satisfy
\begin{align}
    &(\varphi^\dagger)^{{ a\dot{a}}}_{\pm\mp}\cdot \uprho^4 \cdot \uprho^5 \cdot \varphi_{{ b\dot{b}}}^{\pm\mp} = -\frac{i}{4} \delta^{\dot{a}}_{\dot{b}} \left( \sigma^1 \right)^{a}_{b}  \,,
    \qquad
 \quad \quad \; (\varphi^\dagger)^{{ a\dot{a}}}_{\pm\mp}\cdot \uprho^8 \cdot \uprho^4 \cdot \uprho^5 \cdot \varphi_{{ b\dot{b}}}^{\mp\pm} = -\frac{i}{4} \delta^{\dot{a}}_{\dot{b}} \left( \sigma^1 \right)^{a}_{b}  \,, \\
  &(\varphi^\dagger)^{{ a\dot{a}}}_{\pm\mp}\cdot \uprho^3 \cdot \uprho^5 \cdot \varphi_{{ b\dot{b}}}^{\pm\mp} = \frac{i}{4} \delta^{\dot{a}}_{\dot{b}} \left( \sigma^2 \right)^{a}_{b}  \,,
    \qquad
  \quad \quad \; \; \; (\varphi^\dagger)^{{ a\dot{a}}}_{\pm\mp}\cdot \uprho^8 \cdot \uprho^3 \cdot \uprho^5 \cdot \varphi_{{ b\dot{b}}}^{\mp\pm} = \frac{i}{4} \delta^{\dot{a}}_{\dot{b}} \left( \sigma^2 \right)^{a}_{b}  \,, \\
  &(\varphi^\dagger)^{{ a\dot{a}}}_{\pm\mp}\cdot \uprho^3 \cdot \uprho^4 \cdot \varphi_{{ b\dot{b}}}^{\pm\mp} = \frac{i}{4} \delta^{\dot{a}}_{\dot{b}} \left( \sigma^3 \right)^{a}_{b}  \,,
    \qquad
 \quad \quad \; \; \; (\varphi^\dagger)^{{ a\dot{a}}}_{\pm\mp}\cdot \uprho^8 \cdot \uprho^3 \cdot \uprho^4 \cdot \varphi_{{ b\dot{b}}}^{\mp\pm} = \frac{i}{4} \delta^{\dot{a}}_{\dot{b}} \left( \sigma^3 \right)^{a}_{b}  \,, \\
     &(\varphi^\dagger)^{{ a\dot{a}}}_{\pm\mp}\cdot \uprho^7 \cdot \uprho^8 \cdot \varphi_{{ b\dot{b}}}^{\pm\mp} = -\frac{i}{4} \delta^{a}_{b}  \left( \sigma^1 \right)^{\dot{a}}_{\dot{b}} \,,
    \qquad
 \quad \quad \; (\varphi^\dagger)^{{ a\dot{a}}}_{\pm\mp}\cdot \uprho^8 \cdot \uprho^7 \cdot \uprho^8 \cdot \varphi_{{ b\dot{b}}}^{\mp\pm} = -\frac{i}{4} \delta^{a}_{b}  \left( \sigma^1 \right)^{\dot{a}}_{\dot{b}} \,, \\
  &(\varphi^\dagger)^{{ a\dot{a}}}_{\pm\mp}\cdot \uprho^6 \cdot \uprho^8 \cdot \varphi_{{ b\dot{b}}}^{\pm\mp} = \frac{i}{4} \delta^{a}_{b}  \left( \sigma^2 \right)^{\dot{a}}_{\dot{b}} \,,
    \qquad
 \quad \quad \; \;\; (\varphi^\dagger)^{{ a\dot{a}}}_{\pm\mp}\cdot \uprho^8 \cdot \uprho^6 \cdot \uprho^8 \cdot \varphi_{{ b\dot{b}}}^{\mp\pm} = \frac{i}{4} \delta^{a}_{b}  \left( \sigma^2 \right)^{\dot{a}}_{\dot{b}} \,, \\
  &(\varphi^\dagger)^{{ a\dot{a}}}_{\pm\mp}\cdot \uprho^6 \cdot \uprho^7 \cdot \varphi_{{ b\dot{b}}}^{\pm\mp} = \frac{i}{4} \delta^{a}_{b}  \left( \sigma^3 \right)^{\dot{a}}_{\dot{b}} \,,
    \qquad
 \quad \quad \; \; \; (\varphi^\dagger)^{{ a\dot{a}}}_{\pm\mp}\cdot \uprho^8 \cdot \uprho^6 \cdot \uprho^7 \cdot \varphi_{{ b\dot{b}}}^{\mp\pm} = \frac{i}{4} \delta^{a}_{b}  \left( \sigma^3 \right)^{\dot{a}}_{\dot{b}} \,, \\
 &(\varphi^\dagger)^{{ a\dot{a}}}_{\pm\pm}\cdot \uprho^9 \cdot \uprho^4 \cdot \uprho^5 \cdot \varphi_{{ b\dot{b}}}^{\pm\mp} = \pm \frac{i}{4} \delta^{\dot{a}}_{\dot{b}} \left( \sigma^1 \right)^{a}_{b}  \,, 
    \qquad
 \; \; (\varphi^\dagger)^{{ a\dot{a}}}_{\pm\pm}\cdot \uprho^9 \cdot \uprho^7 \cdot \uprho^8 \cdot \varphi_{{ b\dot{b}}}^{\pm\mp} = \pm \frac{i}{4} \delta^{a}_{b}  \left( \sigma^1 \right)^{\dot{a}}_{\dot{b}} \,, \\
 &(\varphi^\dagger)^{{ a\dot{a}}}_{\pm\pm}\cdot \uprho^9 \cdot \uprho^3 \cdot \uprho^5 \cdot \varphi_{{ b\dot{b}}}^{\pm\mp} = \mp \frac{i}{4} \delta^{\dot{a}}_{\dot{b}} \left( \sigma^2 \right)^{a}_{b}  \,, 
    \qquad
 \; \; (\varphi^\dagger)^{{ a\dot{a}}}_{\pm\pm}\cdot \uprho^9 \cdot \uprho^6 \cdot \uprho^8 \cdot \varphi_{{ b\dot{b}}}^{\pm\mp} = \mp \frac{i}{4} \delta^{a}_{b}  \left( \sigma^2 \right)^{\dot{a}}_{\dot{b}} \,, \\
 &(\varphi^\dagger)^{{ a\dot{a}}}_{\pm\pm}\cdot \uprho^9 \cdot \uprho^3 \cdot \uprho^4 \cdot \varphi_{{ b\dot{b}}}^{\pm\mp} = \mp \frac{i}{4} \delta^{\dot{a}}_{\dot{b}} \left( \sigma^3 \right)^{a}_{b}  \,, 
    \qquad
 \; \; (\varphi^\dagger)^{{ a\dot{a}}}_{\pm\pm}\cdot \uprho^9 \cdot \uprho^6 \cdot \uprho^7 \cdot \varphi_{{ b\dot{b}}}^{\pm\mp} = \mp \frac{i}{4} \delta^{a}_{b}  \left( \sigma^3 \right)^{\dot{a}}_{\dot{b}} \,,\\
&(\varphi^\dagger)^{{ a\dot{a}}}_{\pm\pm}\cdot \uprho^9 \cdot \varphi_{{ b\dot{b}}}^{\pm\mp} = \mp \frac{1}{4}\delta^{a}_{b} 
    \delta^{\rm \dot a}_{\rm \dot b}\,,
\end{align}
and 0 otherwise. Defining
\begin{align}
\label{definition phi su2+}
\phi^{a}_{b}  &:= \phi^{\beta_+} \left( \sigma^1 \right)^{a}_{b}  + \phi^{\gamma_+} \left( \sigma^2 \right)^{a}_{b}  - \phi^{\varphi_+} \left( \sigma^3 \right)^{a}_{b}  \\
\label{definition phi su2-}
\phi^{\dot{a}}_{\dot{b}} &:= \phi^{\beta_-} \left( \sigma^1 \right)^{\dot{a}}_{\dot{b}} + \phi^{\gamma_-} \left( \sigma^2 \right)^{\dot{a}}_{\dot{b}} - \phi^{\varphi_-} \left( \sigma^3 \right)^{\dot{a}}_{\rm \dot{b}}
\end{align}
we get that the supersymmetry transformations can be expressed as
\begin{align}
\label{susy massive fermions}
\delta \chi^{ a \dot{a}} &= \frac{1}{2} \left( \slashed{\partial} \phi^{\sf tr} + \frac{\phi^{\sf tr}}{R}  \right) \uptau^3 \kappa^{ a\dot{a}} -\frac{i}{2} \left[ \sin \Omega \left( \slashed{\partial} \phi^{a}_{b}  - \frac{\phi^{a}_{b} }{R} \right) \kappa^{\rm b\dot{a}} + \cos \Omega \left( \slashed{\partial} \phi^{\dot{a}}_{\dot{b}} - \frac{\phi^{\dot{a}}_{\rm \dot{b}}}{R} \right) \kappa^{\rm a\dot{b}} \right] \,, \\
\label{susy massless fermions}
\delta \psi^{ a \dot{a}} &= \frac{1}{2} \slashed{\partial} \phi^{9} \kappa^{ a \dot{a}}+\frac{i}{2} \left( \cos \Omega \; \slashed{\partial} \phi^{a}_{b}  \kappa^{\rm b \dot{a}}-\sin \Omega \; \slashed{\partial} \phi^{\dot{a}}_{\dot{b}} \kappa^{\rm a \dot{b}} \right) \,, \\
\label{susy s1 scalar}
\delta \phi^{9} &=  -\frac{1}{2} \; \overline{ \psi}_{ a \dot{a}}  \kappa^{ a \dot{a}} \,, \\
\label{susy tr scalar}
\delta \phi^{\sf tr} &=  -\frac{1}{2} \; \overline{ \chi}_{ a \dot{a}} \uptau^3 \kappa^{ a \dot{a}} \,, \\
\label{susy s3+ scalars}
\delta \phi^{a}_{b}  &=-\frac{i}{2} \left[ \cos \Omega \left( 2 \; \overline{\psi}_{\rm b\dot{c}} \kappa^{a\dot{c}} - \delta^{a}_{b}  \overline{\psi}_{\rm c\dot{c}} \kappa^{\rm c\dot{c}} \right) -\sin \Omega \left(2 \; \overline{\chi}_{\rm b\dot{c}} \kappa^{\rm a\dot{c}} - \delta^{a}_{b}  \overline{\chi}_{\rm c\dot{c}} \kappa^{\rm c\dot{c}} \right) \right] \,, \\
\label{susy s3- scalars}
\delta \phi^{\dot{a}}_{\dot{b}} &= \frac{i}{2} \left[ \sin \Omega \left( 2 \; \overline{\psi}_{\rm c\dot{b}} \kappa^{\rm c\dot{a}} - \delta^{\dot{a}}_{\dot{b}} \overline{\psi}_{\rm c\dot{c}} \kappa^{\rm c\dot{c}} \right) +\cos \Omega \left(2 \; \overline{\chi}_{\rm c\dot{b}} \kappa^{\rm c\dot{a}} - \delta^{\dot{a}}_{\dot{b}} \overline{\chi}_{\rm c\dot{c}} \kappa^{\rm c\dot{c}} \right) \right] \,.
\end{align}

\end{subsection}

\end{section}

\end{document}